\documentclass[final,3p,times]{elsarticle}

\pdfoutput=1
\UseRawInputEncoding

\usepackage{graphicx}
\usepackage[linesnumbered,ruled,vlined,boxed]{algorithm2e}
\usepackage{multirow}
\usepackage{float}
\usepackage{subfigure}

\begin{document}

\begin{frontmatter}

\title{PHEE: A phased hybrid evaluation-enhanced approach for identifying influential users in social networks}

\author[label1]{Enqiang Zhu}
\author[label1]{Haosen Wang}
\author[label1]{Yu Zhang}
\author[label2]{Kai Zhang}
\author[label3]{Chanjuan Liu \footnote{Corresponding author: chanjuanliu@dlut.edu.cn}}

\address[label1]{Institute of Computing Science and Technology, Guangzhou University, Guangzhou 510006, China.}

\address[label2]{School of
Computer Science and Technology, Wuhan University of Science and Technology, Wuhan, 430065, China.}

\address[label3]{School of Computer Science and Technology, Dalian University of
Technology, China.}

\begin{abstract}
For the purpose of maximizing the spread of influence caused by a certain small number $k$ of nodes in a social network, we are asked to find a $k$-subset of nodes (i.e., a seed set) with the best capacity to influence the nodes not in it. This problem of influence maximization (IM) has wide application, belongs to subset problems, and is $\mathbf{NP}$-hard. To solve it, we should theoretically examine all seed sets and evaluate their influence spreads, which is time-consuming.  Therefore, metaheuristic strategies are generally employed to gain a good seed set within a reasonable time. We observe that many algorithms for the IM problem only adopt a uniform mechanism in the whole solution search process, which lacks a response measure when the algorithm becomes trapped in a local optimum.
To address this issue, we propose a phased hybrid evaluation-enhanced (PHEE) approach for IM, which utilizes two distinct search strategies to enhance the search of optimal solutions: a randomized range division evolutionary (RandRDE) algorithm to improve the solution quality, and a fast convergence strategy.  Our approach is evaluated on 10 real-world social networks of different sizes and types. Experimental results demonstrate that our algorithm is efficient and obtains the best influence spread for  all the datasets compared with three state-of-the-art algorithms, outperforms the time-consuming CELF algorithm on four datasets, and performs worse than CELF on only two networks.
\end{abstract}


\end{frontmatter}

\section{Introduction}\label{introduction}
A social network is defined as a set of nodes connected by one or more relations (such as academic cooperation, shared ideas, and social contacts) \cite{wasserman1994social}.
The last two decades, and especially the period of the COVID-19 pandemic, have seen a growing trend toward analyzing the structure of relationships among users and the flow of communication in  social networks, in areas such as business, health, and sociology \cite{sn3}.
Influence maximization (IM), which aims to identify a certain small set of users (i.e., a seed set) that can maximize the spread of influence in a network, is a topic of major concern in the literature, since it has wide application in the real world \cite{sumith2018}. The best-known  such application is viral marketing \cite{domingos2001mining}, which  utilizes a small number users to propagate a marketing message  through word-of-mouth. To maximize the spread of influence in a network, the initial selected users should have the best ability to influence others, and how to identify these users is naturally formulated as an IM problem. Recent applications include  analyzing human behavior \cite{sn8}, rumor blocking \cite{2015Least}, and personal recommendations \cite{sn3}.

Domingos and Richardson \cite{domingos2001mining} described IM as an algorithm problem in 2001. Research has identified two main challenges when designing such algorithms: how to estimate the spread of influence of a seed set (which has been proved to be $\sharp$\textbf{P}-hard \cite{Kempe2003}), and how to identify a seed set with the maximum influence \cite{r1,r2}.
The spread of influence in a network is generally measured under some fixed diffusion models, including the widely-used    \emph{independent cascade (IC)} and \emph{linear threshold (LT)} \cite{Kempe2003} models.
In 2003, Kempe et al. \cite{Kempe2003} formulated IM as a combinatorial optimization problem, and proved that it is \textbf{NP}-hard under the IC and LT models. Attention has turned to the design of inexact approaches for IM, so that a good (near-optimal) seed set can be found within a reasonable time, and the balance of accuracy of solutions and efficiency of algorithms has become a  central issue in designing a heuristic IM approach.

\subsection {Related work}

An increasing amount of literature has investigated effective and efficient algorithms for the IM problem \cite{tang2015influence,galhotra2016,Seungkeol2017Scalable,sn5,zhu2021local,ai-1}. Given the excellent surveys on IM  \cite{sn7,li2018,survey1}, we give only a brief overview of related work from the aspects of  simulation-based and  metaheuristic approaches.

\subsubsection{Simulation-based approach}

In 2003, in the first systematic study of the IM problem Kempe et al. \cite{Kempe2003} considered both the IC and LT models, and  proposed a simple hill-climbing greedy approach to select the most influential seed set. They proved that the greedy algorithm can guarantee a high approximation ratio ($\approx$ 63\%), which significantly improved the solution accuracy compared with degree-based heuristics and the random-based approach at the time. However, the algorithm requires repeated visits to every node of the network, and  tens of thousands of Monte-Carlo simulations (MCSs) to measure the average influence spreading of a node, and does not apply to large social networks. To improve the efficiency,
Leskovec et al. \cite{kee07} proposed an optimization approach, cost-effective lazy forward (CELF), by leveraging the sub-modular property of the IM problem \cite{Kempe2003}.  Leskovec et al.  proved that CELF can guarantee  an approximation ratio of $\frac{1}{2}(1-\frac{1}{\mathrm{e}})$, which is about 700 times faster than the greedy algorithm of Kempe et al. Nevertheless, CELF still performs poorly in large networks due to the repeated computation of the marginal influence spread of each candidate node. To efficiently handle large networks,
Chen et al. \cite{kee09} proposed a reduction approach by deleting edges,   developed a fast greedy algorithm, NewGreedy, under the IC model, and experimentally showed that it is significantly more efficient than CELF. Not coincidentally,  Goyal et al. \cite{2011CELF} proposed CELF++, which improved the efficiency of CELF. The authors claimed that CELF++ is $35-55\%$ faster than CELF, but  Arora et al. \cite{arora2017} revealed that these two algorithms have almost identical efficiency.

Previous research has established that simulation-based approaches generally have good accuracy, but poor efficiency, which limits their application on large networks. Heuristic search approaches provide a better choice for handling \textbf{NP}-hard problems in large networks,  usually can obtain good (even optimal) solutions within a reasonable time, and have been applied such as the minimum vertex cover \cite{QUAN2021115185}, partition coloring \cite{9241416}, and maximum weight clique   \cite{WANG2020103230} problems. Over the past decade, most research in IM has emphasized metaheuristic approaches to identify the most influential seed set.

\subsubsection{Metaheuristic approaches}

Metaheuristic approaches can generally provide a large range of solution spaces for searching for optimal solutions in a reasonable time, and so reduce the probability of become trapped in a local optimum.   Jiang et al. \cite{jiang11} first used metaheuristic approaches for IM, and used a simulated annealing-based algorithm with an expected diffusion value (a designed measure) to estimate the local influence spread of a seed set and dramatically reduce the running time. However, the method has poor stability in accuracy, possibly due to the restriction of the simulated annealing mechanism and the lack of effective parameter settings  \cite{Zhang2017}. In 2016, Bucur et al. \cite{bucur2016}
proposed a genetic algorithm for IM, and showed that simple genetic operators can help to select nodes with high influence.  Gong et al. \cite{gong2016efficient} considered community-based IM in social networks, and described a memetic algorithm using a community structure and degree centrality to select potential nodes. In 2018,  Cui et al. \cite{Cui2018DDSE} analyzed the drawback (low efficiency) of greedy approaches, proposed a degree-descending search (DDS)  strategy to measure the quality of the seed set, and developed an evolution algorithm, DDSE, with improved efficiency because it does not carry out costly simulations. However, sometimes DDSE performs poorly   because it sets the same upper bound on the size of the candidate set for all networks \cite{Biswas2022}. Singh et al. \cite{singh2020aco}   used a local influence evaluation heuristic (LIEH) with ant colony optimization to approximate local influence.  Zareie et al. \cite{zareie2020identification} adopted an entropy-based metric to estimate the influence spread of nodes, and used a gray wolf optimization algorithm for IM, which outperformed other algorithms in three real-world networks, and had a lower computational time than other metaheuristic methods.
Tang et al. \cite{r2} analyzed the effect of network topology characteristics, based on which
a discrete shuffled frog-leaping algorithm was used for IM. Liu et al. \cite{Qiu2021} used a local-influence-descending (LID) search strategy to construct a candidate set of nodes with relatively large influence, based on which a differential evolution (LIDDE) algorithm was developed for the IM problem. With sufficient iterations,  LIDDE can perform with both accuracy and efficiency, but the results may be affected by network structures. Biswas et al. \cite{biswas2021mcdm,Biswas2022} integrated multi-criteria decision making (MCDM) with some metaheuristics to develop hybrid algorithms for IM.  A hybrid approach that combines MCDM  with simulated annealing was reported in \cite{biswas2021mcdm}, in which it eliminates insignificant nodes (based on MCDM)  at the preliminary phase and uses a modified simulated annealing algorithm  to search for  the optimal solution.  A hybrid two-stage optimization framework that integrates MCDM with an evolution approach was decribed in \cite{Biswas2022}, in which it adopts MCDM  to reduce   computational complexity and uses  a  modified differential evolution (DE) algorithm  to remove insignificant nodes from the candidate set.
 These hybrid algorithms often run effectively, but they require complex designs and architectural knowledge \cite{Biswas2022}.

(Meta)-heuristic-based algorithms usually perform efficiently and have good scalability, but
they cannot guarantee accuracy.
Recent research has suggested that a metaheuristic algorithm does not exist that can outperform simulation-based greedy algorithms on a wide range of instances, and almost all previous approaches may perform poorly in either efficiency or accuracy \cite{Biswas2022}. Therefore, how to balance between solution quality and running time of IM algorithms is still a challenge. In addition, we observe that few heuristic IM algorithms adopt a distinct mechanism in the whole search process to enhance the quality of the solution, which may weaken the ability to
escape local optima. Hence, we propose a phased hybrid evaluation-enhanced (PHEE) approach to solve the IM problem in social networks, with a randomized range division evolutionary (RandRDE) algorithm to improve the quality and   diversity of solutions, and  an accelerated convergence strategy to efficiently search for optimal solutions.
 Experimental results on a range of instances demonstrate that PHEE outperforms state-of-the-art algorithms in both accuracy and efficiency in almost all   instances.

\subsection{Contributions}

This paper makes the following contributions:

 (a) The PHEE framework, integrating an evolutionary approach with accelerated convergence, is proposed to solve the IM problem, adopting distinct vertex influence estimation strategies to enhance the quality of solutions in the search process;

 (b) A random range search mechanism  enhances the diversity of solutions;

 (c) Two PHEE-based algorithms for the IM problem  employ different heuristic vertex influence estimation strategies;

 (d) Numerical experiments on 10 datasets demonstrate the effectiveness and performance of the proposed algorithm.

The remainder of this paper is organized as follows. Section  \ref{set2} provides preliminary information,  including notation,  the definition of the influence maximization problem, and the diffusion model adopted in this paper. PHEE  is presented in section \ref{sec3}. Section \ref{sec4}  discusses our evaluation algorithm, $RandRDE()$, and the accelerated convergence strategy for solution generation is presented in section  \ref{sec5}. Section \ref{sec6} is devoted to the design and analysis of experiments, and section \ref{sec7} provides conclusions and discusses future work.

\section{Preliminaries}\label{set2}
Social networks often are modeled as graphs with vertices and edges respectively representing
users and their social relationships. We refer to social networks as ``graphs", and follow the standard terminologies in graph theory. We describe some necessary notation, and formally define influence maximization and our adopted diffusion model.

\subsection{Notation}\label{set2-1}

Only simple graphs are considered in this paper. A graph $G$ with vertex set $V$ and edge set $E$ can be denoted by $G=(V,E)$. We use $V(G)$ and $E(G)$ to denote the vertex set and edge set, respectively, of $G$. For any pair of vertices $u,v\in V$, we say that they are \emph{adjacent} (or \emph{neighbors}) in $G$ if there exists an edge $e\in E$ connecting them, and denote by $uv$ the edge $e$, i.e., $e=uv$, where $u$ and $v$ are called the  \emph{endpoints} of $e$. An edge is \emph{incident} with a vertex that is an endpoint of the edge. The \emph{degree} of a vertex $v\in V$, denoted by $d_G(v)$, is the number of edges incident with $v$ in $G$. Let $N_G(v)=\{u| u\in V$ and $uv\in E\}$  be the set of neighbors of $v$ in $G$, and let $N_G[v]=N_G(v)\cup \{v\}$.  For any $S\subseteq V$, let $N_G(S)=\cup_{v\in S} N_G(v)$, and denote by $G[S]$  the \emph{subgraph of $G$ induced by $S$}, i.e., the graph obtained from $G$ by deleting all vertices in $V\setminus  S$ and their incident edges.

\subsection{Influence maximization problem}\label{set2-2}

We define the \textbf{influence maximization} (\textbf{IM}) problem. A graph $G=(V,E)$ (a social network) has active and inactive vertices, where active vertices can influence their inactive neighbors in some way. As a result, an inactive vertex  becomes active if it is successively influenced by its active neighbors; otherwise, it remains inactive. The dynamic process of vertices from inactive to active represents influence propagation, and the initial set of active vertices is called a \textbf{seed set} of the graph. Given a seed set $S$, its \emph{influence} $\sigma(S)$ is defined as the  number of active vertices at the end of the influence propagation process (under a specific diffusion model). The IM problem refers to finding a seed set of $k$  vertices with the maximum influence, where $k$ is a given small positive integer. Formally, the IM problem can be described as follows:

\emph{Input}:  Graph $G$, positive integer $k$, and diffusion model $\sigma$

\emph{Goal}: Find a seed set $S^*$ of cardinality $k$  with maximum influence $\sigma(S^*)$, i.e.,

\begin{equation}\label{equ1}
  S^* = \arg\max\limits_{S\subseteq V, |S|=k} \sigma(S),
\end{equation}
where the maximum is taken over all seed sets $S$ of cardinality $k$.

\subsection{Influence diffusion models}\label{sec3}
As mentioned above, influence can spread with the aid of some diffusion models. Several models exist for the measurement of influence propagation in networks \cite{sumith2018}.  We adopt the independent cascade (IC) model, due to its simplicity, to estimate influence propagation for seed sets.

The simplest of the dynamic cascade models,  IC was first mentioned by Goldenber et al. \cite{2001Talk}, and then was used by Kempe et al. \cite{Kempe2003}.  The model supposes that every vertex is either active or inactive, and activation occurs in discrete time-steps (i.e., an inactive neighbor can be activated by each of its active neighbors), which can be described as follows. Denote by $u$ a vertex that is activated at step $t$, and let $v$ be an arbitrary inactive neighbor of $u$. Then $u$ has only one chance to activate $v$  with \emph{active probability} $p_{u,v}$, which is predefined for each edge $uv$ that is independent of all previous attempts to activate $v$. If $u$ succeeds, then $v$ becomes active at step $t+1$; otherwise, $v$ is still inactive, and $v$ will never be activated by $u$ in subsequent steps. If an inactive vertex $v$ has more than one newly activated neighbor at step $t$, it can be activated by them one by one, in any order. The diffusion process stops when  no more vertices can be activated.

\section{Phased hybrid evaluation-enhanced framework} \label{sec3}

PHEE involves two search strategies with totally different characteristics  in two distinct search stages: one explores potential vertices, and the other accelerates convergence.

Among inexact approaches to  the IM problem, metaheuristic approaches have attracted attention for their problem independence and generality \cite{Cui2018DDSE,Biswas2022}.
Designing a metaheuristic algorithm for an \textbf{NP}-hard problem does not require an investigation of the properties of the problem, which saves considerable time and energy over problem-specific algorithms \cite{NG2018104,Biswas2022}, such as the greedy-based CELF \cite{kee07}.
Evolutionary approaches have been reported to have  good global search capability \cite{Cui2018DDSE,Qiu2021,Biswas2022},  and adaptive simulated annealing approaches can accelerate convergence \cite{biswas2021,jiang11,WU2017576,alsalibi2022comprehensive}. Hence, we adopt these two approaches.

Algorithm \ref{alo-1} has four steps,   starting with sub-procedure $SortV()$, which  sorts vertices in non-increasing order based on a certain estimation criterion (lines 2 and 3), whose results are received by $RandRDE()$ (Algorithm \ref{alo-randRDE}), which outputs a  candidate set of vertices (lines 4 and 5).  $RandRDE()$ is an evolutionary algorithm, which uses a random range division strategy to improve the diversity of solutions, utilizing a range-specified function to partition the search range of vertices when generating random solutions. We restrict the number of iterations in $RandRDE()$ to reduce computation in this stage. Next, $ConstructSS()$ (Algorithm \ref{alo-initial})  constructs an initial solution by an independent approach, so as to disturb the current solution space, and a self-adaptive simulated annealing algorithm, $AdapSAA()$ (Algorithm \ref{alo-asa}), searches the optimal solutions.

\begin{algorithm}[htp!]
\footnotesize
\caption{A Phased Hybrid Evaluation-Enhanced (PHEE) Framework for IM}\label{alo-1} {\em }

\KwIn{a  graph $G$, a positive integer $k$}
\KwOut{A seed set $S$ of cardinality $k$}

\Begin{
\textbf{Step 1: Vertex sorting}

\vspace{0.1cm}

$SVet \leftarrow SotV()$;// $SotV()$ returns a sorted list of vertices in non-increasing order in terms of some certain estimation criterion

\vspace{0.1cm}

\textbf{Step 2: On the basis of SVet, construct candidate pool \emph{CSSet}  by evolution approach}

\vspace{0.1cm}

$CSSet \leftarrow RandRDE()$; // $RandRDE()$ is an evolution algorithm that returns a  candidate set $CSSet$ of vertices

\vspace{0.1cm}

\textbf{Step 3: Construct initial solution using an approach independent of the above two steps}

\vspace{0.1cm}

$IniS \leftarrow ConstructSS()$; // $ConstructSS()$ is an independent approach that construct an initial solution (a seed set);

\textbf{Step 4: Search global optimal solution $S$ by adaptive simulated annealing approach}

$S \leftarrow  AdapSAA()$; // $AdapSAA()$ is an adaptive simulated annealing approach based on Steps 2 and 3

\Return $S$;

}
\end{algorithm}

Previous  algorithms for the IM problem used various heuristic strategies  in the initial stage to improve the accuracy and diversity of estimating the influence of vertices, or random operators that can cover a wider range of potential vertices in the solution search. Our framework employs two random search strategies in independent stages, which has at least two advantages. First, it improves the ability to escape from local optima using an independent solution construction method and adaptive simulated annealing when $sortV()$ and $RandRED()$ cannot generate high-quality solutions. Second, it utilizes the advantages of different vertex influence estimation methods and solution search strategies to enhance the reliability of the solution. In addition, we observe that our framework can be used on other \textbf{NP}-hard problems.

\section{Random Range Division Evolution} \label{sec4}

We describe $RandRDE()$ (Algorithm \ref{alo-randRDE}), which is based on a sorted list of vertices.

\subsection{Vertex sorting} \label{vertex-sorting}

PHEE first estimates the influence spread of each vertex, and sorts them accordingly. The measurement criteria of vertex influence have been extensively investigated. Centrality-based measures, such as degree, eigenvector, closeness, and betweenness centrality \cite{abbasi2012betweenness,malliaros2016locating}, are popular because a few vertices may play a vital role in the spread of influence, and these usually have high centrality \cite{biswas2021mcdm}. While these criteria are effective, they cannot guarantee accuracy on networks with different structures. Therefore, studies have been conducted on the  partition of network structures. One such decomposition is based on the $k$-shell (initially the $k$-core \cite{Seidman1983}), and has been used to identify influential vertices  in networks in which influence originates in a single vertex \cite{kitsak2010identification}. Zeng et al. \cite{zeng2013ranking} observed that $k$-shell decomposition ignores the influence of removed vertices when estimating the spread of influence of vertices, and proposed a modified $k$-shell decomposition, mixed degree decomposition (MDD),  based on the  concepts of the \emph{residual degree} and \emph{exhausted degree}. Another modification, the gravity centrality index (GCI), was inspired by the gravity formula \cite{ma2016identifying}.
Considering the good performance of MDD and GCI \cite{maji2020systematic}, we use them to rank vertices at the first step of PHEE.

\subsubsection{Mixed degree decomposition}

We recall the definitions of $k$-shell ($k$-core) and $k$-shell decomposition. A \emph{$k$-shell} of a graph $G=(V,E)$ is a subgraph $H_k=G[V']$ induced by $V'\subseteq V$ such that $d_{H_k}(v)\geq k$ for every $v\in V'$, and every vertex $v'\in V\setminus V'$ is adjacent to at most $k-1$ vertices of $V'$, where $k$ is a positive number less than $|V|$. The \emph{shell number} $\mathcal{S}(u)$ of a vertex $u\in V$ is the maximum $k$ such that $u$ belongs to a $k$-shell of $G$. The \emph{shell decomposition} of a graph $G$ aims to partition its structure based on the $k$-shell, i.e., to partition vertex set $V$ by shell numbers.
The shell number of vertices is calculated by recursively selecting a vertex of minimum degree, deleting it and its incident edges, and updating its neighbors' degrees \cite{batagelj2003m}. Although  vertices can be efficiently ranked based on the shell number, the information of removed vertices is ignored, which may affect the accuracy of vertex influence estimation. Therefore, the method of mixed degree decomposition is proposed, which involves  the residual degree and exhausted degree.

Let $G=(V,E)$ be a graph, and $D\subseteq V$ an arbitrary subset of vertices. The \emph{residual degree} (\emph{exhausted degree}) of a vertex  $v\in V\setminus D$ with respect to $D$, denoted by $k^r_D(v)$ ($k^e_D(v)$), is defined as the number of neighbors of $v$ in $V\setminus D$ ($D$), i.e., $k^r_D(v)$= $d_{G[V\setminus D]}(v)$ and $k^e_D(v)$= $d_{G[D\cup \{v\}]}(v)$, and its \emph{mixed degree}   with respect to $D$ is
\begin{equation}\label{mdd}
k^m_D(v)= k^r_D(v) + \lambda \times k^e_D(v),
\end{equation}
where the weight coefficient $\lambda$ is a tunable real number in [0,1]. \emph{Mixed degree decomposition} (MDD) is a generalization of $k$-shell decomposition,  which can be roughly described as recursively deleting vertices of minimum mixed degree, and updating the set of removed vertices and the mixed degrees of all remaining vertices. We rank vertices based on MDD \cite{zeng2013ranking}, as shown in Algorithm \ref{alo-mdd}.

Algorithm \ref{alo-mdd} initializes every element in $Rank_{mdd}[]$ with value 1,  the ordered set $DelV$ with the empty set, the mixed degree $k^m_{DelV}(v)$ of each vertex $v\in V(G)$ with degree $v$, and  $min_{mdd}$ with the minimum mixed degree,  where $Rank_{mdd}[]$ is an array that contains the rank of each vertex, and $DelV$ is the set of removed vertices (lines 2--5). A loop iteratively ranks vertices (lines 6--14). In each iteration, if a vertex $v$ exists in the current graph such that $k^m_{DelV}(v)\leq min_{mdd}$,  let $R$ be the set of all such vertices, i.e., $R=\{v |k^m_{DelV}(v)\leq min_{mdd}\}$. Then, we define the rank of  vertices $v\in R$ as $min_{mdd}$, add vertices of $R$ to $DelV$, update the mixed degree of vertices in $N_G(R)$ (with respect to the new $DelV$), and delete vertices in $R$ and their incident edges from $G$ (lines 8--12); otherwise, we increase the value of $min_{mdd}$ (line 14).  The algorithm returns the reverse of $DelV$, i.e., a sorted sequence in nonincreasing order of the rank of vertices (line 15).

\begin{algorithm}[htp!]
\footnotesize
\caption{$SortV_{mdd}$}\label{alo-mdd} {\em }

\KwIn{a  graph $G=(V,E)$, a real number $\lambda$}
\KwOut{A sorted list $SVet$ of vertices}

\Begin{

$Rank_{mdd}[]\leftarrow 1$, $DelV \leftarrow \emptyset$; //$Rank_{mdd}[]$ is an array that contains the rank of each vertex, and $DelV$ represents the ordered set of removed vertices

\For{each $v\in V$}{$k^m_{DelV}(v) \leftarrow d_G(v)$; }

$min_{mdd} \leftarrow \min\{k^m_{DelV}(v)|v\in V\}$;

\vspace{0.1cm}

\While{$|DelV|<|V|$}{

\If{there exists $v\in V(G)$ with $k^m_{DelV}(v)\leq min_{mdd}$} {

$R \leftarrow$ the set of vertices $v$ with $k^m_{DelV}(v)\leq min_{mdd}$;

\vspace{0.1cm}

$Rank_{mdd}[v] \leftarrow min_{mdd}$ for each $v\in R$;

\vspace{0.1cm}

$DelV \leftarrow DelV \cup R$;

\vspace{0.1cm}

Update the mixed degree of vertices in $N_G(R)$; // $N_G(R)=\cup_{v\in R} N_G(v)$

\vspace{0.1cm}

$G \leftarrow G-R$; // $G-R$ is the resulting graph obtained from $G$ by deleting vertices in $DelV$ and its incident edges;
}

\Else{
Increase the value of $min_{mdd}$ by $++min_{mdd}$;
}

}

$SVet \leftarrow $ the reverse of $DelV$;

\Return $SVet$;
}
\end{algorithm}

\subsubsection{Gravity centrality index}

Observing that the influence of a vertex may increase if it has high-influence neighbors, and the spread of influence from one vertex to another may decrease when the distance between them increases, Ma et al. \cite{ma2016identifying} introduced the gravity centrality index (GCI)  based on the shell number to estimate the influence of vertices, which is inspired by the classical gravity formula of Isaac Newton. Given a graph $G=(V,E)$, to estimate the influence of a vertex  by the gravity formula, the shell number of the vertex is viewed as its mass, and the shortest path between two vertices   as their distance. Hence, the gravity centrality index of a vertex $v$ is defined as

\begin{equation}\label{gci}
\mathcal{G}(v)=\sum\limits_{u\in N^{r}_G(v)} \frac{\mathcal{S}(v) \times \mathcal{S}(u) }{(d_G(v,u))^2},
\end{equation}
where $d_G(v,u)$ is the distance between $v$ and $u$, $N^{r}_G(v)$ is the set of vertices $u$ with $d_G(v,u)\leq r$ (a given value), and  $\mathcal{S}(v)$ is the shell number of $v$. It is known that the shell number can be calculated in $O(m)$ time where $m$ is the number of edges \cite{batagelj2003m}. Therefore, a sorted list $SVet$ of vertices in nonincreasing order based on the GCI of vertices can be determined efficiently for small $r$.

$SVet$ can be constructed to reduce the search space for the IM problem, and   should possess the property (to some extent) that previous vertices in $SVet$ are more likely to be selected as top-$k$ influential vertices.

\subsection{Random range division evolution}

Evidence suggests that although the search space for the IM problem is large, only a small number of vertices actively participate in the spreading process \cite{biswas2021mcdm}. Hence, strategies have been proposed to reduce the search space \cite{Cui2018DDSE,singh2020aco,biswas2021mcdm}. We observe that little research has considered reducing the solution space according to the quality of solutions generated in the search, which may limit the diversity of the solution space. We propose an evolutionary algorithm, $RandRDE()$, whose pseudocode is shown in  Algorithm \ref{alo-randRDE}, to improve the diversity of  PHEE.

The algorithm constructs an initial population $\mathbf{X}$  by  calling subprocedure $InitialPop()$ (see Algorithm \ref{alo-initialpop}), where $\mathbf{X}$ can be viewed as a group of individuals, each a seed set (i.e., a set of $k$ vertices).
A loop (limited by iteration number $g_{max}$) iteratively optimizes the population $\mathbf{X}$; its three components are mutations (Algorithm \ref{alo-mut}), crossover (Algorithm \ref{alo-cro}), and selection (Algorithm \ref{alo-sel}) (lines 3--8). The algorithm returns a candidate set $CSSet$ of vertices (lines 9--12).
A random range division (RRD) strategy for the subprocedures (including $InitialPop()$,  $RDEMutation()$, and $RDECrossover()$) can flexibly expand the coverage of the solution search according to the sizes of graphs, to guarantee the diversity and the influence quality of vertices in the candidate set.

\begin{algorithm}[htp!]
\footnotesize
\caption{$RandRDE$}\label{alo-randRDE} {\em }

\KwIn{A graph $G=(V,E)$, the size $k$ of seed set, an ordered list $SVet$ of vertices, the population size $pop$, the mutation probability $mp$, the crossover probability $cp$, and the number of iterations $g_{max}$}

\vspace{0.1cm}

\KwOut{A candidate set $CSSet$; // $CSSet$ contains sorted vertices  according to  the number of occurrences of the vertex $v$ in the population individuals}

\Begin{

$\mathbf{X} \leftarrow InitialPop(G,SVet,  div\_factor, pop, k)$;

\vspace{0.1cm}

\While{$iteration_{num} \leq  g_{max}$}{

$\mathbf{X}_{m} \leftarrow RDEMutation(\mathbf{X}, SVet, mp, pop, k)$;

\vspace{0.1cm}

$\mathbf{X}_{c} \leftarrow RDECrossover(G,  SVet, \mathbf{X}, \mathbf{X}_{m}, cp, pop, k)$;

\vspace{0.1cm}

$\mathbf{X} \leftarrow RDESlection(\mathbf{X}, \mathbf{X}_{cro},  pop)$;

\vspace{0.1cm}

$iteration_{num} \leftarrow iteration_{num}+1$;

}

$CSSet \leftarrow $ the set of vertices that appear in $\mathbf{X}$; // i.e., the union of all individual seed sets in the final population  $\mathbf{X}$;

\vspace{0.1cm}

\Return $CSSet$;
}
\end{algorithm}

\subsubsection{Random range division strategy}
RRD randomly disturbs the candidate pool each time the algorithm constructs an individual. Whenever an individual is to be generated, RRD randomly generates an upper bound $ub$ and constructs an exclusive candidate pool containing the first $ub$ vertices in the ordered list $SVet$ (obtained on the basis of MDD or GCI; see section  \ref{vertex-sorting}) for this individual. Thus, different individuals may be generated from different candidate pools, which improves the diversity of the final candidate set built by $RandRDE()$. The upper bound $ub$ should satisfy the property that the ratio $\frac{ub}{n}$ is not too large (for efficiency) when the number of vertices $n$ is relatively large, and not too small when $n$ is small (for accuracy). Based on the above considerations, the upper bound is determined as

\begin{equation}\label{upbound}
up\_bound(k,n,p)=k+n(\frac{k}{n-k})^{1-p}\cdot \sin(\frac{\pi p}{2}),
\end{equation}
where $k$ is the cardinality of the seed set,  and $p$  is the \emph{random division parameter}, a probability between $0.1$ and $0.5$.

It is clear that $up\_bound(k,n,p)> k$. Whenever a seed set is  constructed, the procedure will randomly generate a real number for $up\_bound()$, which determines an upper bound $ub$ for constructing a candidate pool (i.e., the first $ub$ vertices of $VSet$). According to equation (\ref{upbound}), the cardinality of the candidate pool (i.e., $ub$) is associated with not only the seed set size $k$ and graph size $n$, but with the random division parameter $p$. So, the candidate pools for individuals may be distinct. The RRD strategy is applied throughout the $RandRDE()$ algorithm.

\subsubsection{$RandRDE()$ initialization}\label{ini-sol}

We denote by $\mathbf{X}=\{\mathbf{X}(1),\ldots, \mathbf{X}(pop)\}$ a population of size $pop$ for the IM problem, where $\mathbf{X}(i)$ for $i=1,2,\ldots, pop$ is the $i$th individual (a seed set of size $k$) of $\mathbf{X}$. A poor-quality initial population may delay convergence. Therefore, an RRD-based  method is introduced to diversify the initial population  based on the top-$k$ vertices. Algorithm \ref{alo-initialpop} describes this approach. The construction is based on an ordered list $VSet$ of vertices. As explained earlier, it is assumed that vertices in the front rank of $VSet$ may have high influence. So, population construction starts with an initialization of each $\mathbf{X}(i)$ by the set of first $k$ vertices in $VSet$ (line 5). RRD constructs a candidate pool $VSet_{i}$ for each individual $\mathbf{X}(i)$ (lines 6--8). To diversify the initial population, a user-defined diversity factor $div\_factor$ is used to probabilistically determine  whether a vertex in an individual is replaced by another, according to a random value in the range [0,1) (lines 10--12). Note that to guarantee that the final individual $\mathbf{X}(i)$ for each $i\in \{1,2,\ldots, pop\}$ contains $k$ distinct vertices,  the vertex replacing $v$ should not belong to the current individual $\mathbf{X}(i)$.

\begin{algorithm}[htp!]
\footnotesize
\caption{$InitialPop(G, SVet, div\_factor, pop,k)$}\label{alo-initialpop} {\em }

\KwIn{A graph $G=(V,E)$, the diversity factor $div\_factor$, an ordered list $SVet$ of vertices, the population size $pop$, the size $k$ of seed set}

\vspace{0.1cm}

\KwOut{A population $\mathbf{X}$ that contains $pop$ individuals}

\Begin{

$\mathbf{X}  \leftarrow \emptyset$;

$SVet_{topk} \leftarrow$ the set containing the first $k$ vertices of $SVet$;

\For{each $i\leq pop$}{
$\mathbf{X}(i) \leftarrow SVet_{topk}$;

$p \leftarrow $ generating a random value in the range [0.1,0.5];

\vspace{0.1cm}

$ub \leftarrow  up\_bound(k, |V|, p)$;

$SVet_{i} \leftarrow $  the set containing the first $up$ vertices of $SVet$;

\vspace{0.1cm}

\For{each vertex $v \in \mathbf{X}(i)$}{

$ran_{num} \leftarrow $ a random value in the range [0,1);

\If{$ran_{num}<div\_factor$}{

Replace $v$ with a vertex in $SVet_{i}\setminus \mathbf{X}(i)$ randomly;
}

}

Add $\mathbf{X}(i)$ into $\mathbf{X}$;

}

\Return $\mathbf{X}$;
}
\end{algorithm}

\subsubsection{RandRDE mutation}\label{mutation}
To avoid being trapped in a local optimum, a mutation operator for RandRDE further expands the range of solutions. The mutation rule is analogous to that in initialization (Algorithm \ref{alo-initialpop}), as described in Algorithm \ref{alo-mut}. Based on an initial population $\mathbf{X}$, $RDEMutation$ uses the RRD strategy again to modify each individual $\mathbf{X}(i)$ of $\mathbf{X}$, by introducing a different diversity factor $mp$ (i.e., mutation probability) from $div\_factor$ (lines 4--10).

\begin{algorithm}[htp!]
\footnotesize
\caption{$RDEMutation(\mathbf{X}, SVet, mp, pop,k)$}\label{alo-mut} {\em }

\KwIn{An initial population $\mathbf{X}$,   an ordered list $SVet$ of vertices, the mutation probability $mp$, the population size $pop$, the size $k$ of seed set}

\vspace{0.1cm}

\KwOut{A mutation population $\mathbf{X}_{m}$ of size $pop$}

\Begin{

$\mathbf{X}_{m}  \leftarrow  \mathbf{X}$;

\For{each $i\leq pop$}{

$p \leftarrow $ generating a random value in the range [0.1,0.5];

$ub \leftarrow  up\_bound(k, |V|, p)$;

$SVet_{i} \leftarrow $  the set containing the first $up$ vertices of $SVet$;

\vspace{0.1cm}

\For{each vertex $v \in \mathbf{X}(i)$}{

$ran_{num} \leftarrow $ a random value in the range [0,1);

\If{$ran_{num}<mp$}{

Replace $v$ with a vertex in $SVet_{i}\setminus \mathbf{X}(i)$ randomly;
}

}

}

\Return $\mathbf{X}_{m}$;
}
\end{algorithm}

\subsubsection{RandRDE crossover}\label{crossover}

Evolutionary algorithms often employ a crossover strategy to generate a mixed population based on the initial and mutated populations, to increase the probability of finding a global optimal solution. A common rule for generating a crossover individual is to select vertices from an initial individual and its corresponding mutated individual based on a predefined  probability $cp$, a random number $ran_{num}$, and the circumstance of the current vertex \cite{Qiu2021}. Specifically, let $\mathbf{X}$ and $\mathbf{X}_{m}$ be the generated initial population and mutated population, respectively, of size $pop$. Let $k$ be the size of the seed set, i.e., the cardinality of an individual.  To determine the $j$th vertex $\mathbf{X}_c(i,j)$  of the $i$th crossover individual $\mathbf{X}_c(i)$, for $i\in \{1,\ldots, pop\}$ and $j\in \{1,\ldots, k\}$, the algorithm considers five conditions:

 (\textbf{C1}) $ran_{num}<cp$ $\wedge$ $\mathbf{X}_{m}(i,j)\notin \mathbf{X}_{c}(i)$;

 (\textbf{C2}) $ran_{num}<cp$ $\wedge$ $\mathbf{X}_{m}(i,j)\in \mathbf{X}_{c}(i)$ $\wedge$ $\mathbf{X}(i,j)\notin \mathbf{X}_{c}(i)$;

  (\textbf{C3}) $ran_{num}\geq cp$ $\wedge$ $\mathbf{X}(i,j)\notin \mathbf{X}_{c}(i)$;

  (\textbf{C4}) $ran_{num}\geq cp$ $\wedge$ $\mathbf{X}(i,j)\in \mathbf{X}_{c}(i)$ $\wedge$ $\mathbf{X}_{m}(i,j)\notin \mathbf{X}_{c}(i)$;

  (\textbf{C5})  $\mathbf{X}(i,j)\in \mathbf{X}_{c}(i)$ $\wedge$ $\mathbf{X}_{m}(i,j)\in \mathbf{X}_{c}(i)$.
Then, based on our RRD strategy, $\mathbf{X}_c(i,j)$ can be generated as

\begin{equation}\label{croequ}
{\small
\mathbf{X}_{c}(i,j)=\left\{\begin{array}{lll}
\vspace{0.1cm}
\mathbf{X}_{m}(i,j), ~if ~ \textbf{C1}~holds\\
\vspace{0.1cm}
\mathbf{X}(i,j), ~if ~ \textbf{C2}~holds \\
\vspace{0.1cm}
\mathbf{X}(i,j), ~if ~ \textbf{C3}~holds\\
\vspace{0.1cm}
\mathbf{X}_m(i,j), ~if ~ \textbf{C4}~holds \\
\vspace{0.1cm}
v\in SVet_i\setminus \mathbf{X}_c(i), ~if ~ \textbf{C5}~holds \\
\end{array},\right.
}
\end{equation}
where $SVet_i$ is the candidate pool for $\mathbf{X}_{c}(i,j)$ that is built according to the RRD strategy.
During the  generation of $\mathbf{X}_{c}(i,j)$, if the random number $ran_{num}$ is less than $cr$, the selection of $\mathbf{X}_{c}(i,j)$ prioritizes vertex $\mathbf{X}_{m}(i,j)$, and otherwise vertex $\mathbf{X}(i,j)$. In particular, if both $\mathbf{X}_{m}(i,j)$ and $\mathbf{X}(i,j)$ have been filled in $\mathbf{X}_{c}(i)$ before $\mathbf{X}_{c}(i,j)$; then, the algorithm randomly selects a vertex in $SVet_i\setminus \mathbf{X}_c(i)$ for
$\mathbf{X}_{c}(i,j)$. See Algorithm \ref{alo-cro} for details.

\begin{algorithm}[htp!]
\footnotesize
\caption{$RDECrossover(\mathbf{X}, \mathbf{X}_m, SVet, cp, pop,k)$}\label{alo-cro} {\em }

\KwIn{An initial population $\mathbf{X}$, a mutated population $\mathbf{X}_m$,  an ordered list $SVet$ of vertices, the crossover probability $cp$, the size of population  $pop$, the size of seed set $k$}

\vspace{0.1cm}

\KwOut{A crossover population $\mathbf{X}_{c}$ of size $pop$}

\Begin{

$\mathbf{X}_{c} \leftarrow  \emptyset$;

\For{each $i\leq pop$}{

\For{each $j\leq k$} {

$p \leftarrow $ generating a random value in the range [0.1,0.5];

$ub \leftarrow  up\_bound(k, |V|, p)$;

$SVet_{i} \leftarrow $  the set containing the first $ub$ vertices of $SVet$;

$ran_{num} \leftarrow $ a random value in the range [0,1);

update $\mathbf{X}_{c}(i,j)$ by Equation \ref{croequ};

}

}

\Return $\mathbf{X}_{c}$;
}
\end{algorithm}

Algorithm \ref{alo-cro} initializes the crossover population $\mathbf{X}_{c}$ with an empty set (line 2) and obtains each crossover individual  $\mathbf{X}_{c}(i)$ by mixing the corresponding mutated individual $\mathbf{X}_{m}(i)$ with its
original individual $\mathbf{X}(i)$ according to equation \ref{croequ} (lines 3--9).

\subsubsection{RandRDE selection}\label{selection}

In the final selection step, the algorithm estimates the influence of the individuals  in the initial and crossover populations, and selects the influential ones as  the final individuals.  The accuracy of the adopted estimation method always determines the quality of solutions. Considering this, RandRDE employs the EDV strategy \cite{jiang11}, which measures the quality of a seed set $S$ by counting the number of vertices in $N_G(S)$ that are inactivated by $S$ with probability $p$. EDV can be formulated as

\begin{equation}\label{edv}
EDV(S) = k + \sum\limits_{v\in N_G(S)\setminus S} (1- (1-p)^{\tau(v)}),
\end{equation}
where $k$ is the size of the seed set, $p$ is the active probability in the IC model, and $\tau(v)=|N_G(v)\cap S|$ is the number of vertices in $S$ that are adjacent to $v$ in graph $G$.

Then the algorithm uses $EDV$ to update the initial population $\mathbf{X}$  by the crossover population $\mathbf{X}_c$. To this end, the EDVs of  each individual $\mathbf{X}(i)$ in $\mathbf{X}$ and its corresponding  crossover individual $\mathbf{X}_c(i)$ are calculated, and the individual with the larger EDV is selected to update  $\mathbf{X}(i)$. Clearly, the influence of the initial individuals can be improved after this selection process.  Algorithm \ref{alo-sel} describes the process.

\begin{algorithm}[htp!]
\footnotesize
\caption{$RDESelection( G, \mathbf{X}, \mathbf{X}_c, pop,k)$}\label{alo-sel} {\em }

\KwIn{A graph $G=(V,E)$, an initial population $\mathbf{X}$, a crossover  population $\mathbf{X}_c$,  the size of population  $pop$, the size of seed set $k$}

\vspace{0.1cm}

\KwOut{A selection population $\mathbf{X}_s$}

\Begin{

$\mathbf{X}_{s} \leftarrow  \emptyset$;

\For{each $i\leq pop$}{

$\mathbf{X}_{s}(i) \leftarrow  \arg\max \{ EDV(\mathbf{X}(i)), EDV(\mathbf{X}_c(i))\}$;

}

\Return $\mathbf{X}_{s}$;
}
\end{algorithm}

\section{Fast convergence strategy of solution generation } \label{sec5}

\begin{algorithm}[H]
\footnotesize
\caption{$AdapSAA(G, CSSet,  k, T_i, T_f, \theta, N)$}\label{alo-asa} {\em }

\KwIn{A graph $G=(V,E)$, a candidate set $CSSet$ of vertices,
 the size of seed set $k$, initial temperature $T_i$ , final
temperature $T_f$, cooling coefficient $\theta$, the number $N$ of neighborhood moves at each temperature ($T_t$) level}

\vspace{0.1cm}

\KwOut{A seed set $S^*$ of cardinality $k$}

\Begin{

$S^* \leftarrow  ConstructSS(G, k)$;

$T_t \leftarrow T_i, t\leftarrow 0, r\leftarrow 0$;

\vspace{0.1cm}

\While{$T_t>T_f$}{

\vspace{0.1cm}

\For{$i<N$}{
$u \leftarrow $ select an arbitrary vertex from $S^*$;

\vspace{0.1cm}

$v \leftarrow $ select an arbitrary vertex from $CSSet\setminus S^*$;

\vspace{0.1cm}

$S \leftarrow (S^*\setminus \{u\})\cup \{v\}$;

\vspace{0.1cm}

\If{$EDV(S)>EDV(S^*)$}{

\vspace{0.1cm}
$S \leftarrow S^*$;

\vspace{0.1cm}

$r\leftarrow 0$;

\vspace{0.1cm}

}\Else{ $r \leftarrow  r+1$; }

}

$T_{t+1} \leftarrow T_t - \theta \ln(r+1)$;

\vspace{0.1cm}

$t \leftarrow t+1$;

}

\Return $S^*$;
}
\end{algorithm}

\begin{algorithm}[H]
\footnotesize
\caption{$ConstructSS(G, k)$}\label{alo-initial} {\em }

\KwIn{A graph $G=(V,E)$, the size of seed set $k$}

\vspace{0.1cm}

\KwOut{A seed set $IniS$ of cardinality $k$}

\Begin{

$IniS \leftarrow \emptyset$;

\vspace{0.1cm}

\While{$|IniS|<k$}{

$v \leftarrow $ a vertex in $G$ with the maximum degree;

\vspace{0.1cm}

$IniS \leftarrow IniS\cup \{v\};$

\vspace{0.1cm}

$G \leftarrow  G-v$; // deleting $v$ and its incident edges from $G$
}

\Return $Inis$;
}
\end{algorithm}

Generally, after generations of evolution, an algorithm will select an individual (solution)  of good quality (according to some estimation function, e.g., EDV) from the final population, and utilize a simple optimization-local
search to improve it. However, the quality of such a solution is too dependent on the estimation strategy of  vertex influence and the evolutionary search strategy, which may affect the stability of the algorithm.  We break this routine by regarding  $RandRED()$ as the first strategy search of PHEE, aiming to construct a candidate set of potential vertices based on their influence. To improve the efficiency of PHEE, we end the iteration of $RandRED()$ when its search efficiency drops, and utilize a different   strategy to complete the next-stage solution search based on the candidate set obtained by $RandRED()$. The new search strategy should converge quickly. To this end, adaptive simulated annealing \cite{biswas2021} is employed due to its effectiveness and efficiency. Given that both MDD and GCI (used to estimate the influence spread of vertices) are based on $k$-shell, which may ignore the overlapping influence among vertices, we use an independent set-based method (Algorithm \ref{alo-initial}) to construct the initial solution for adaptive simulated annealing (Algorithm \ref{alo-asa}).

Algorithm \ref{alo-asa} starts with the construction of an initial solution $S^*$ (line 2) by using a greedy idea of recursively finding a vertex of the maximum degree in the current graph (Algorithm \ref{alo-initial}; lines 4-5). Whenever a vertex is determined, the graph is  updated by deleting the vertex and its incident edges (Algorithm \ref{alo-initial}; line 2). This initially constructed $S^*$ is clearly an independent set (i.e., it does not contain any pair of adjacent vertices). Therefore, the overlapping influence among $S^*$ is reduced to some degree. Next, at each temperature level $T_t$, it carries out an iteration to improve the current optimal solution $S^*$ using EDV estimation, where a variable $r$ (initialized as 0) is used to record the number of consecutive unsuccessful improvements (lines 5--13). The algorithm utilizes an adaptive  temperature control function (\cite{biswas2021}) that can control the temperature drop according to the search progress; this leads to the balance between the accuracy and quality of solutions (line 14).

\section{Experiments} \label{sec6}
We experimentally compared  the performance of our algorithm    with four baseline algorithms.

\subsection{Datasets}

We evaluated PHEE on 10 real-world social networks of different sizes and types, including four academic collaboration networks, two email exchange networks, two who-trust-whom online social networks, a decentralized
peer-to-peer file sharing network, and a particular users network, whose statistics are shown in Table \ref{tab-dataset}, where  $|V|, |E|, d_{ave}$, and Type represent the number of vertices, number of edges, average degree, and type of graph ($U$: undirected;  $D$: directed), respectively.  Netscience \cite{tang2019identification} is a co-authoship network that describes academic collaborations between scientists in network theory and experiments. Email-un \cite{tang2019identification} is an email exchange network between universities. CA-GrQc \cite{singh2019lapso}, CA-HepTh \cite{tang2019identification}, and CA-AstroPh \cite{r2} are co-authoship networks describing academic collaborations between authors from the General Relativity section of Arxiv, the High Energy Physics section of Arxiv, and the  Astrophysics section of Arxiv.
Gnutella \cite{singh2019lapso} is a decentralized peer-to-peer file sharing network from Nullsoft's Gnutella; soc-Epinions1 \cite{r2} and soc-Epinions2 \cite{rossi2015network} are who-trust-whom online social networks of customer review site Epinions.com. Slashdot \cite{r2} is a social network that describes the sharing of technology-related news between users. Email-Eu \cite{r2} is a large email exchange network from a European research institute. These datasets can be downloaded from NETWORK REPOSITORY (https://networkrepository.com/index.php).

\begin{table} [H]
\centering
\footnotesize
\caption{Information on ten social networks}
\vspace{0.1cm}
\renewcommand\tabcolsep{10.0pt}
\renewcommand{\arraystretch}{1.2}
\label{tab-dataset}
\begin{tabular}{c c c c c c}
\hline
    Datasets & $|V|$ & $|E|$ & $d_{ave}$  & Type & Reference\\
    \hline
    Net-science& 1589&	2742&	3.45&		U & \cite{tang2019identification}\\
    Emial-un&	1133&	5451&	9.622&		D &   \cite{tang2019identification} \\
    CA-GrQc&	5242&	14495&	5.53&		U &  \cite{singh2019lapso}\\
    CA-HepTh&	15233&	58891&	7.73&		U &  \cite{tang2019identification} \\
    CA-AstroPh&	18772&	198110&	21.107&		U &  \cite{r2}\\
    Gnutella&	62586&	147892&	4.73&		D & \cite{singh2019lapso}\\
    soc-Epinions1&	75888&	508837&	6.71&		D & \cite{r2} \\
    soc-Epinions2&	26594&	100126&	7.53&		D & \cite{rossi2015network}\\
    Slashdot&	77360&	905468&	23.409&		U & \cite{r2} \\
    Eaiml-eu&	265214&	420045&	3.168&		U& \cite{r2} \\
    \hline
\end{tabular}
 \end{table}

\subsection{Baseline algorithms}

We compare PHEE  with  four baseline algorithms for the IM problem. Since  PHEE involves two metaheuristic approaches, evolution and simulated annealing,  the state-of-the-art algorithms selected for comparison are mainly of the same type.

\begin{enumerate}

\item [(1)] Cost-effective lazy forward (CELF) \cite{kee07}  is a greedy-based approach  that requires at least 10,000 Monte Carlo simulations to estimate the marginal influence of a vertex at each
round. It can generally obtain a good solution, but it is time-consuming. Therefore, it is often used as a reference to show that a proposed algorithm can efficiently obtain a high-quality  solution.

\item [(2)] Degree-descending search evolution (DDSE) \cite{Cui2018DDSE} is a swarm intelligence-based algorithm  that uses a  DDS strategy to estimate   influence spread.

\item [(3)] Simple additive weighting adaptive simulated annealing (SAW-ASA) \cite{biswas2021} is a multi-criteria decision making and simulated annealing-based metaheuristic algorithm that integrates four centrality-based strategies and a modified version of simulated annealing.

\item [(4)] Local-influence-descending differential evolution (LIDDE) \cite{Qiu2021} is a differential evolution algorithm that utilizes  an objective function, the expected diffusion impact value (EDIV), to improve the accuracy of the algorithm.

\end{enumerate}

In our experiments, parameter values in the above algorithms were set according to the corresponding papers.

To  demonstrate the effectiveness of  PHEE for the IM problem, we used the MDD and GCI strategies to estimate the significance of vertices at the first step of PHEE, generating corresponding algorithms MDD-PHEE and GCI-PHEE.

\subsection{Experiments for the parameters of our algorithm}

To fully explore the advantages of the evolutionary approach and simulated annealing in our algorithm and best trade off between the quality of the seed set and computational efficiency, we conducted five experiments to tune the parameter settings, including the weight coefficient $\lambda$ in MDD (equation \ref{mdd}), the maximum number of iterations $gmax$ in $RandRDE()$ (algorithm \ref{alo-randRDE}), the range of  random division parameter $p$ in RRD (equation \ref{upbound}), and the mutation probability $mp$ and   crossover probability $cp$ in $RandRDE()$. For convenience, we use $N_i$, $i=1,2,\ldots,10$, to  represent  	Net-science,	Email-un,	CA-GrQc,	CA-HepTh,	CA-AstroPh,	Gnutella,	soc-Epinions1,	Slashdot,	soc-Epinions2,	Email-eu, respectively.

\subsubsection{Weight coefficient $\lambda$ in MDD}

We conducted  sensitivity experiments on datasets $N_i, i=1,2,\ldots, 10$, with values of $\lambda$ from 0.1 to 1.0,  increasing by 0.1. Let $\lambda_i=\frac{i}{10}$, $i=1,2,\ldots, 10$. To compare the results obtained by different $\lambda$ values, we considered the influence spread (under PHEE) of each case, and used the Friedman statistical test \cite{friedman1937use} for comparison. For $N_i$ and $\lambda_j$, $i,j=1,2,\ldots, 10$, we   ran $MDD-PHEE$ over 10 seed set sizes (from 10 to 100), and recorded their influence spreads. On each network, we ranked $\lambda_i$, $i=1,2,\ldots,10$, according to the influence spreads for each seed set size, and calculated the mean ranks for each, as shown in Table \ref{lambda}. The overall rank of  each $\lambda_i$ was determined by the average of all its mean ranks on the 10 networks; see the last column in Table \ref{lambda}. Given that IM is a maximization problem,  we used a higher mean rank value to reflect an algorithm with a better influence spread.
From the results in Table \ref{lambda},  we can see that $\sigma_7$ is the maximum. Therefore, we set $\lambda$ at 0.7 in our algorithm.

\begin{table} [htp!]
\centering
\footnotesize
\caption{Friedman test results for ranking different $\lambda$ values.}
\label{lambda}
\renewcommand\tabcolsep{10.0pt}
\renewcommand{\arraystretch}{1.1}
\begin{tabular}{c c c c c c c c c c c c}
\hline
     $\lambda$ & 	$N_1$ & 	$N_2$ & 	$N_3$ & 	$N_4$ & 	$N_5$& 	$N_6$  & 	$N_7$ & 	$N_8$ & 	$N_9$ & 	$N_{10}$& 	Overall rank \\
    \hline
 $\lambda_1$	& 1.6	& 3.5&	1.5	&1.6&	1	& 5.6&	5.4	& 4.5	&1.2&	6&	3.19\\
$\lambda_2$ 	&4.9&	4.1&	4.2&	3.6&	3.5&	6.2&	3.9&	5.2	&4.7&	4.8	&4.51\\
$\lambda_3$&	2.95	&4.4&	3.3	&3.1&	2.9&	4.9&	5.15&	5.9&	3.3&	5.2&	4.11\\
$\lambda_4$&	4.55&	7.1	&3.7&	4.2&	7.6&	3.6&	6.1&	6.1&	5.9&	5.5&	5.435\\
$\lambda_5$&	6.2&	3.2&	4.7&	5.3&	7.5&	6.3&	6.3&	6.2&	7.9&	5.1&	5.87\\
$\lambda_6$ &	7.1&	5.4&	6.8&	7.1&	6.9&	5.6&	6.2&	4.5&	5.9&	5.8&	6.13\\
$\lambda_7$&	6.25&	6.2&	6.9&	6.9&	7.7&	7.1&	4&	7.05&	7&	8.1&	6.72\\
$\lambda_8$&	6.7&	7.4&	8.2&	7.8&	5.6&	5.8&	6.6&	7.15&	6.9&	5&	6.715\\
$\lambda_9$&	7.75&	7.8&	8&	7.7&	5.5&	5&	5.55&	4.6&	6.9&	6&	6.48\\
$\lambda_{10}$ &	7&	5.9	& 7.7&	7.7&	6.8&	4.9&	5.8&	3.8	& 5.3&	3.5&	5.84\\
    \hline
\end{tabular}
 \end{table}

\subsubsection{Maximum number of iterations $gmax$ for $RandRDE$}
\begin{figure}[H]
  \centering
  \subfigure[$gmax$]{
\includegraphics[width=8cm]{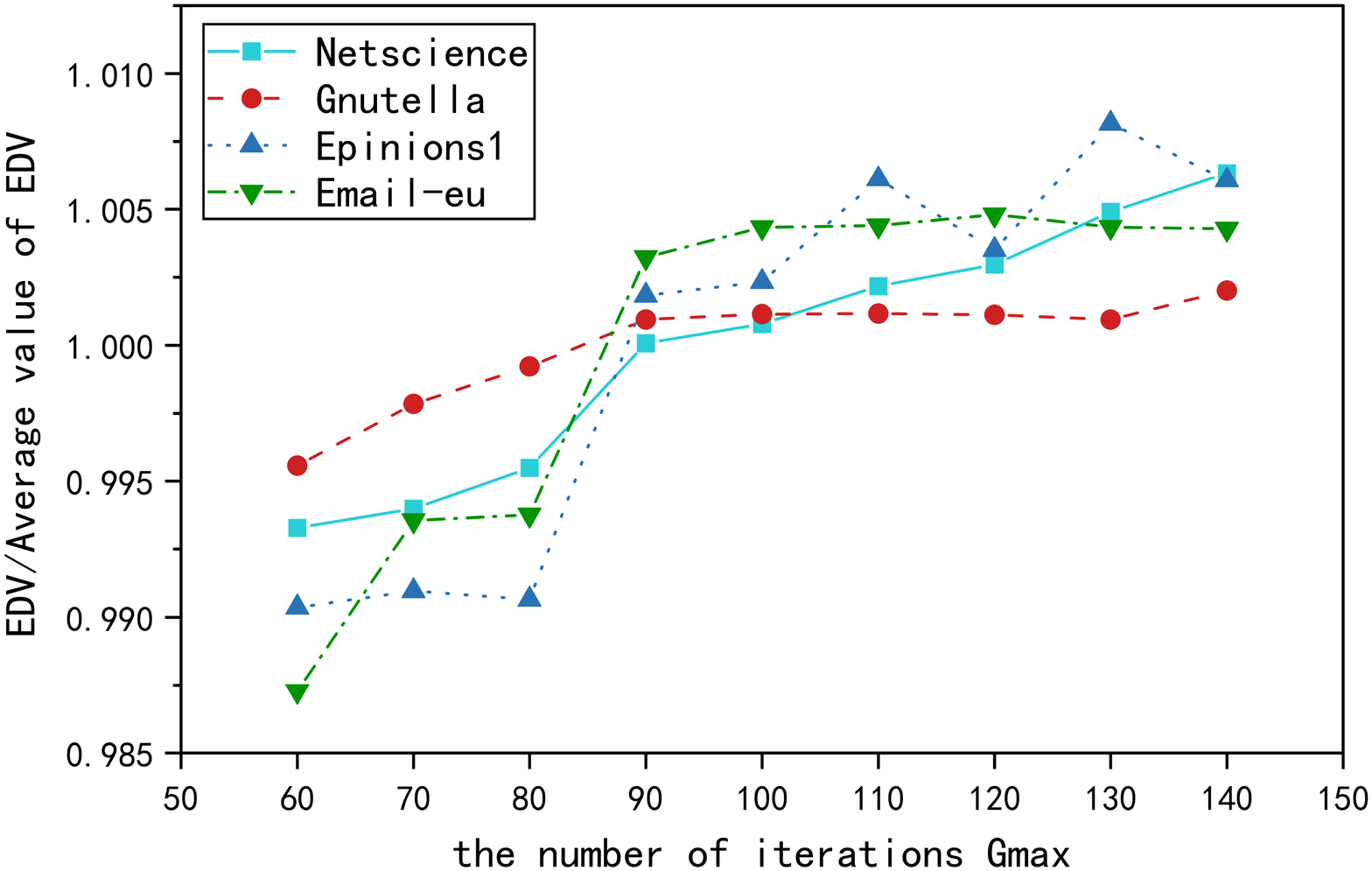}
}
 \subfigure[$p$]{
\includegraphics[width=8cm]{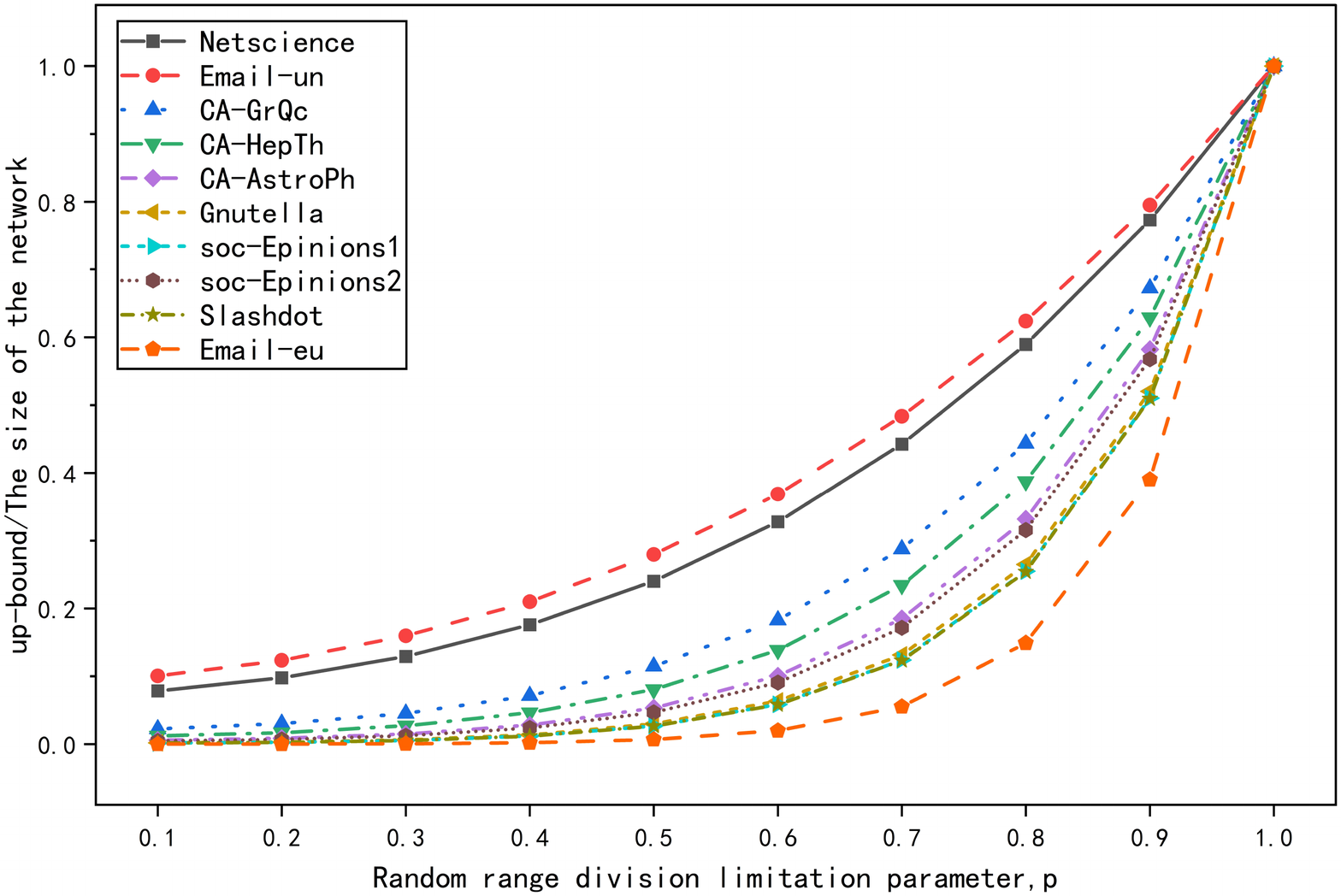}
}
\caption{Sensitivity experiments for the maximum number of iterations  $gmax$ for RandRDE and the range of random division parameter $p$}\label{add-1}
\end{figure}

Since PHEE is a phased approach that utilizes two distinct search strategies to generate the final seed set, we must constrain  the iteration number $gmax$ of $RandRDE()$, which saves as much time as possible for the simulated annealing search to obtain the final optimal solution. To select the best $gmax$ value, we conducted experiments to investigate the quality of solutions and computational cost under values of $gmax$ from $60$ to $140$, increasing by 10. We estimated the results by the $EDV$ strategy. Let  $g_{j}=j$ for $j=60, 70, \ldots, 140$. For each pair ($N_i$, $g_j$), $i= 1,2,\ldots, 10$ and $j \in \{60, 70, \ldots, 140\}$,  it first calculated the $EDV$ value of all seed sets  in the population obtained by $RandRDE()$ (where the size of the seed set is 100), and selected the maximum $EDV$ value as the representative of ($N_i$, $g_j$), denoted by $edv_{i,j}$. For each $i\in \{1,2,\ldots, 10\}$ and $j\in \{60, 70, \ldots, 140\}$, we used
 $$
\gamma_{i,j}= \frac{ 9\times edv_{i,j}}{\sum_{\ell\in \{60, 70, \ldots, 140\}} edv_{i,\ell} }
$$
as the fitness value of  ($N_i$, $g_j$), and assumed that a greater $\gamma_{i,j}$ indicates a better-quality solution under $g_j$. Figure \ref{add-1} (a) shows the experimental results on four networks selected according to size: Netscience (small), Gnutella (medium), and  soc-Epinions1 and Email-eu (large). From the results, we see that a better solution can be obtained when $gmax = 100$,  and more iterations do not greatly improve its quality. Therefore, to save computational cost, we set $gmax$ to 100.

\subsubsection{Range of  random division parameter  $p$ }

The aim of RRD (equation \ref{upbound}) is to construct search spaces   for different networks. Note that a small search space will result in a poor diversity of solutions,  while a large search space will reduce the search efficiency.  Both may affect the quality of the solution. We carried out experiments to compare the search space under different values of $p$ (from 0.1 to 1.0, increasing by 0.1), with $k$ set to 100.

As shown in Figure \ref{add-1} (b), with $p$ values ranging from 0.1 to 0.3, the search space is relatively small, consisting of  only  vertices with high rank, and can ensure a seed set whose quality is not so bad. With $p$ values of 0.4 and 0.5, the search space expands gradually, which can  cover a sufficient range of vertices with high influence. When $p$ is from 0.6 to 0.8, the search space is too large, which  possibly generates a poor  random solution, while increasing the computational cost. When $p$ is 0.9 or 1.0, the search space covers almost the whole network. Considering the trade-off between the quality and diversity  of solutions,
we set the range of this parameter from 0.1 to 0.5.

\subsubsection{Mutation probability
$mp$}

\begin{figure}[H]
  \centering
  \subfigure[mp]{
\includegraphics[width=8cm]{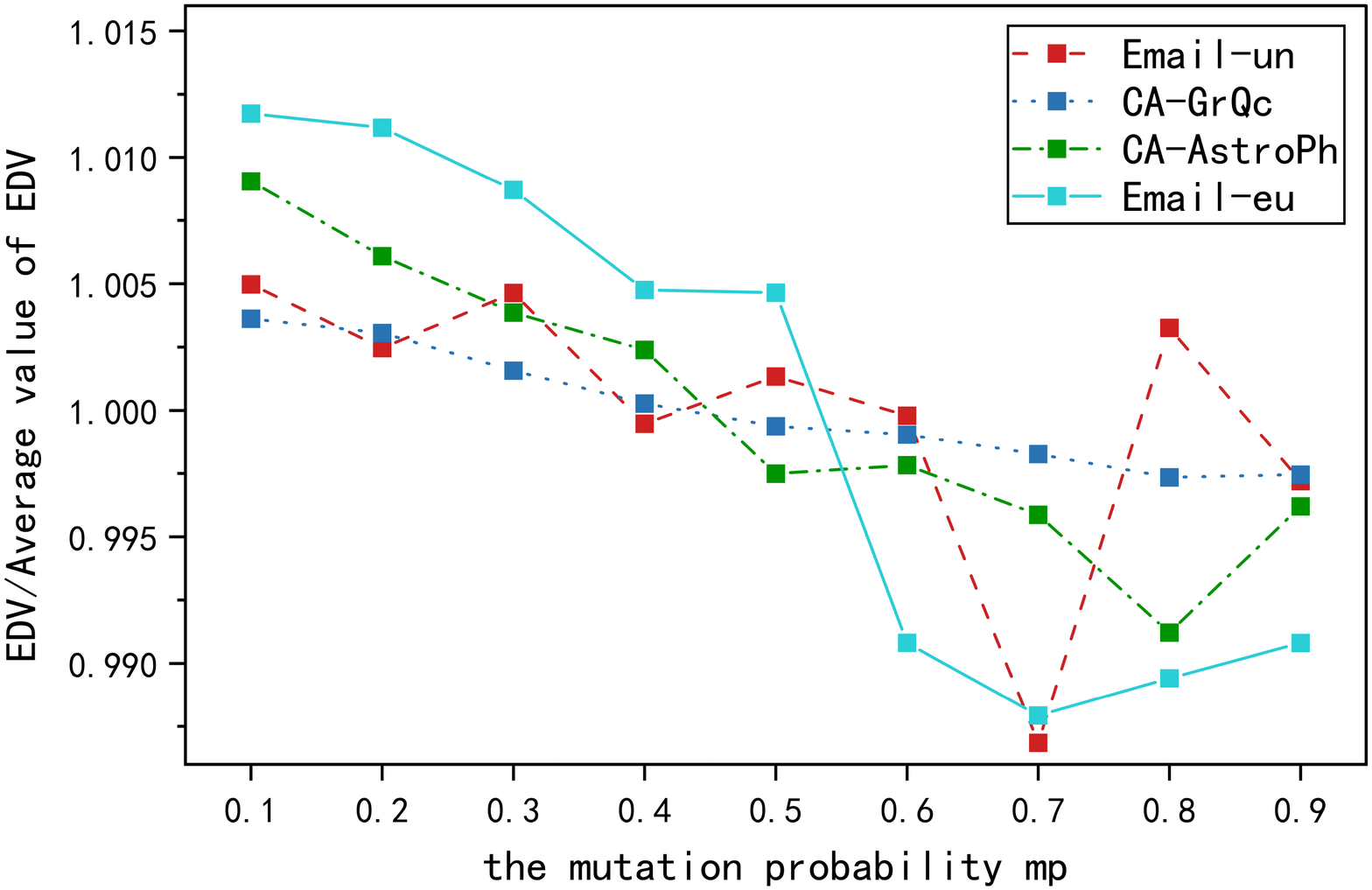}
}
 \subfigure[cp]{
\includegraphics[width=8cm]{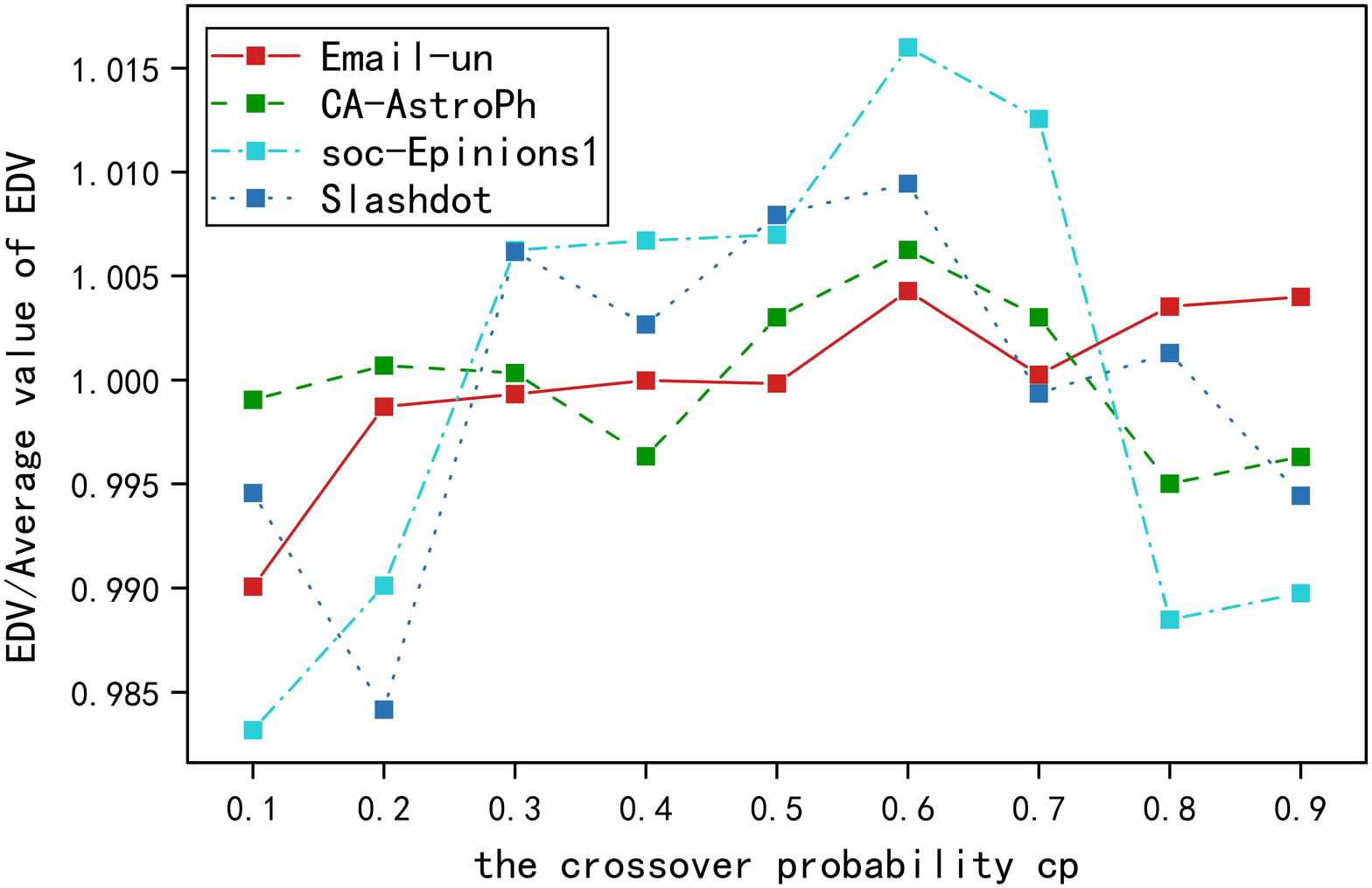}
}
  \caption{The variations of  fitness  values with different $mp$ and $cp$ values}\label{mp-para}
\end{figure}

To investigate the impact of different $mp$ values, we experimentally compared the performance of our proposed algorithm with $mp$ values from 0.1 to 0.9 (increasing by 0.1).
For each network $N_i$, we estimated the results by the $EDV$ strategy. Let  $m_{j}=\frac{j}{10}$ for $j=1, 2, \ldots, 9$. For each pair ($N_i$, $g_j$), $i= 1,2,\ldots, 10$ and $j \in \{1,2, \ldots, 9\}$,  it calculates $EDV$ for all seed sets  in the population obtained by $RandRDE()$ (here, the size of the seed sets is 100) and selects the maximum $EDV$ value to represent ($N_i$, $g_j$), denoted by $edvm_{i,j}$. Then, for each $i\in \{1,2,\ldots, 10\}$ and $j\in \{1,2, \ldots, 9\}$, we use
 $$
\beta_{i,j}= \frac{ 9\times edv_{i,j}}{\sum_{\ell\in \{1,2,\ldots, 9\}} edvm_{i,\ell} }
$$
as the fitness value of  ($N_i$, $m_j$), and assume that a greater $\beta_{i,j}$ indicates a better-quality of solution under $m_j$. See Figure \ref{mp-para} (a) for the experimental results on four networks selected according to the sizes of networks: small (Email-un and CA-GrQc), medium (CA-AstroPh), and large (Email-eu). From the results, we see that the EDV fitness values are best when mp = 0.1, and they descend with the increase of $mp$ for almost all these networks. Therefore, we set $mp$ as 0.1.

\subsubsection{Crossover probability $cp$}

To explore the impact of $RandRDE()$ with different $cp$ values, we conducted the same experiment as   for mutation probability
$mp$.
Figure \ref{mp-para} (b) shows the variations of  the fitness  values with different $cp$ values, which we can see   generally trend upward   when $cp$ increases in the interval [0.1, 0.6], and   reach a peak when $cp=0.6$, but decrease when $cp$ increases in the interval [0.6, 0.9]. Therefore, we
set the crossover probability $cp$ as 0.6.

Notice that in the above experiments, parameters other than the tested parameter were set as follows: $\lambda=0.7$ (weight coefficient in MDD), $pop$=10 (number of populations in $RandRDE$), $gmax=100$ (maximum number of iterations in $RandRDE$), $div\_factor = 0.6$ (diversity factor $div_factor$ of initial population), $mp$=0.1 (mutation probability), $cp=0.6$ (crossover probability), $T_i=2000$ (initial temperature in $ADapSAA$), $T_f=10$ (final
temperature in $ADapSAA$), $N=15$ (number of neighborhood moves at each temperature level, and $\theta = 5$ (cooling coefficient).

\subsection{Experiment setup}

All of the experimental simulations were implemented under an IC diffusion model with active diffusion probabilities $ap$ (we assume that all edges $uv$ in one network have the same $ap$ value), varying with the size and structure of networks.   $ap$ was set to $0.05$ for NetScience and Email networks (due to their sparsity and small size), and $0.01$ for the other networks. The size of the seed set was set to 10, 20, 30, 40, 50, 60, 70, 80, 90, and 100.

All experiments were implemented in C++, compiled by g++, and run on a Linux server with an Intel Xeon Gold 6254 CPU @ 3.10 GHz, with 251 GB RAM under CentOS 7.5.

To obtain accurate estimates of the influence spread of a candidate seed vertex for CELF, the number
of MCSs was set to 10,000 runs (based on \cite{Kempe2003},\cite{kee07}), and the average was taken. For  comparison, the influence spreads
for the other algorithms were calculated by setting 1000 runs, following the usual practice \cite{Cui2018DDSE,tang2019identification,biswas2021}.

\subsection{Comparison on influence spread} \label{com-influenc}
The influence spread of the seed set obtained by an algorithm for the IM problem is the main criterion by which to estimate its performance, where to produce a seed set with a larger influence spread is regarded as better performance. We ran all of the compared algorithms with seed set sizes  from 10 to 100 on the 10 networks under the IC model.

\begin{figure}[H]
\begin{center}
\subfigure[Netscience]{
\includegraphics[width=7cm]{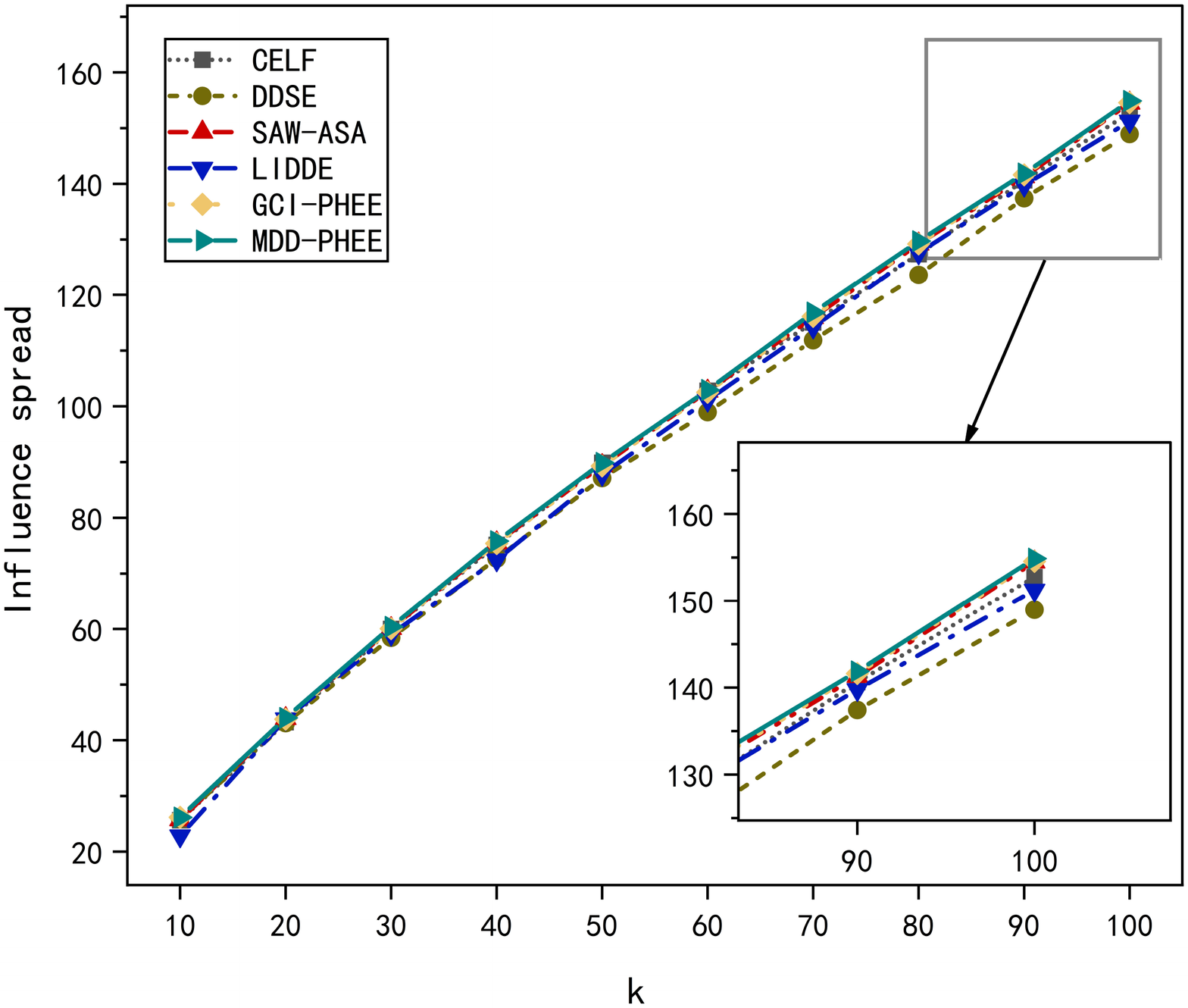}
}
\subfigure[Email-un]{
\includegraphics[width=7cm]{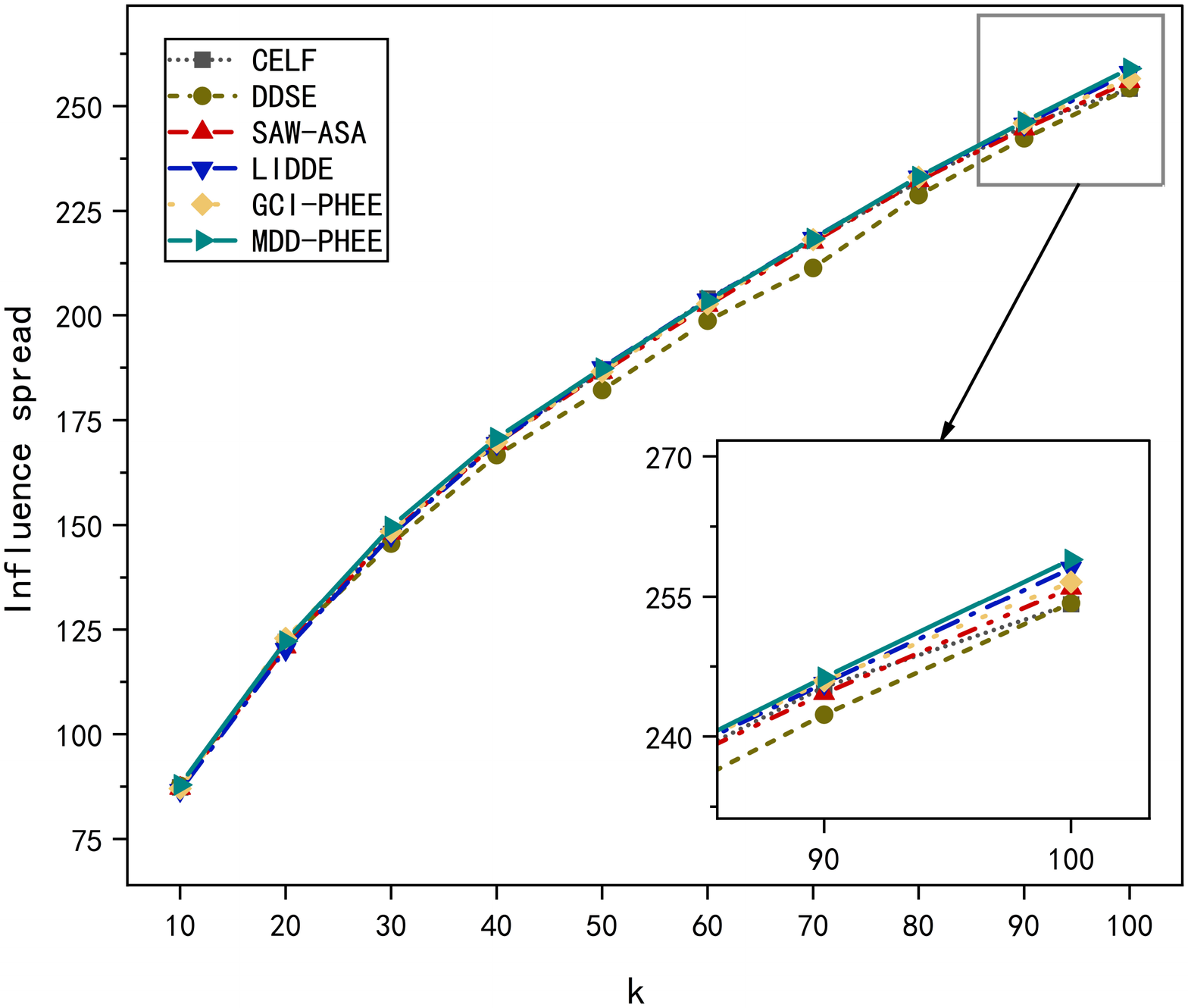}
}
\subfigure[CA-GrQc]{
\includegraphics[width=7cm]{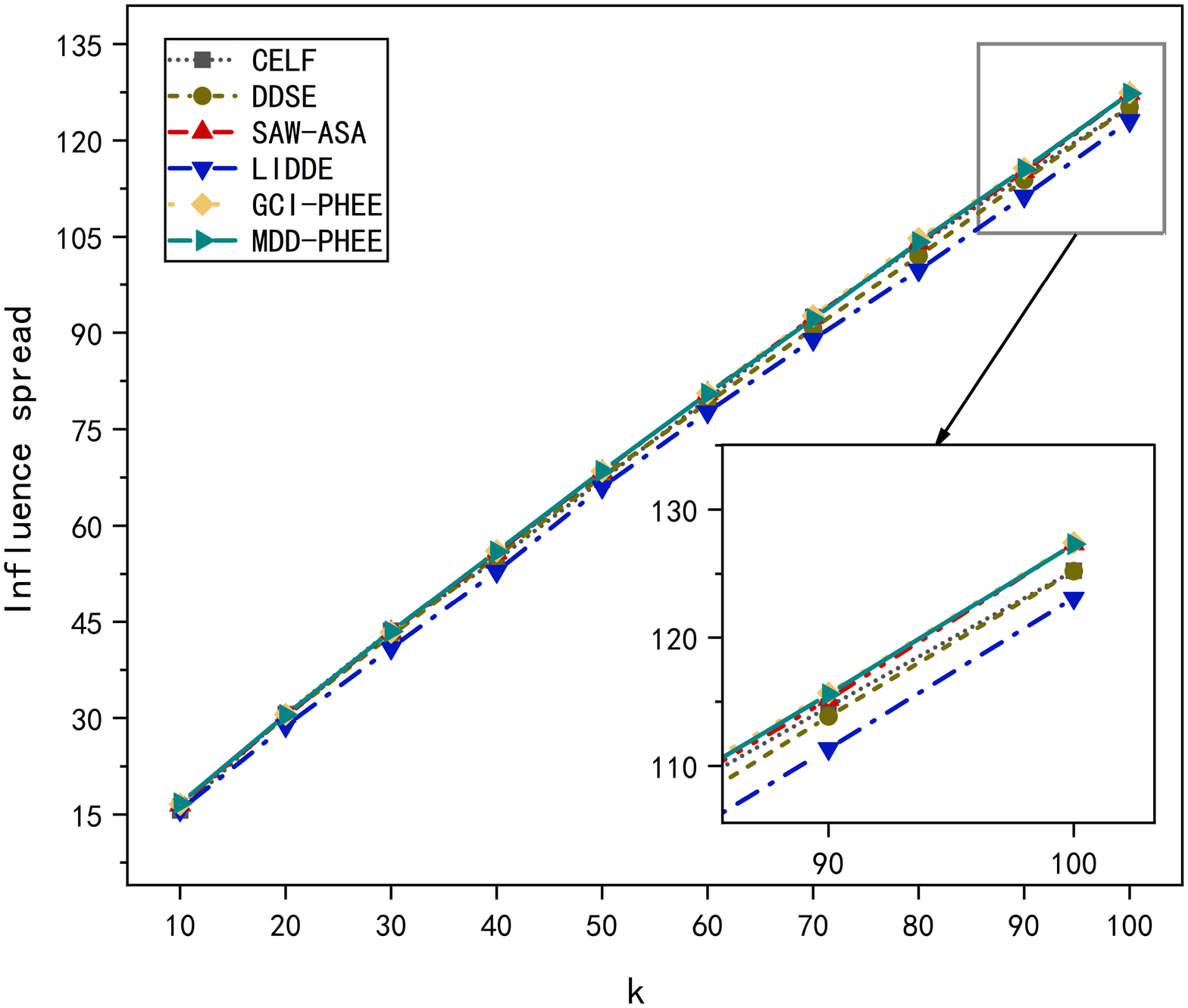}
}
\subfigure[CA-HepTh]{
\includegraphics[width=7cm]{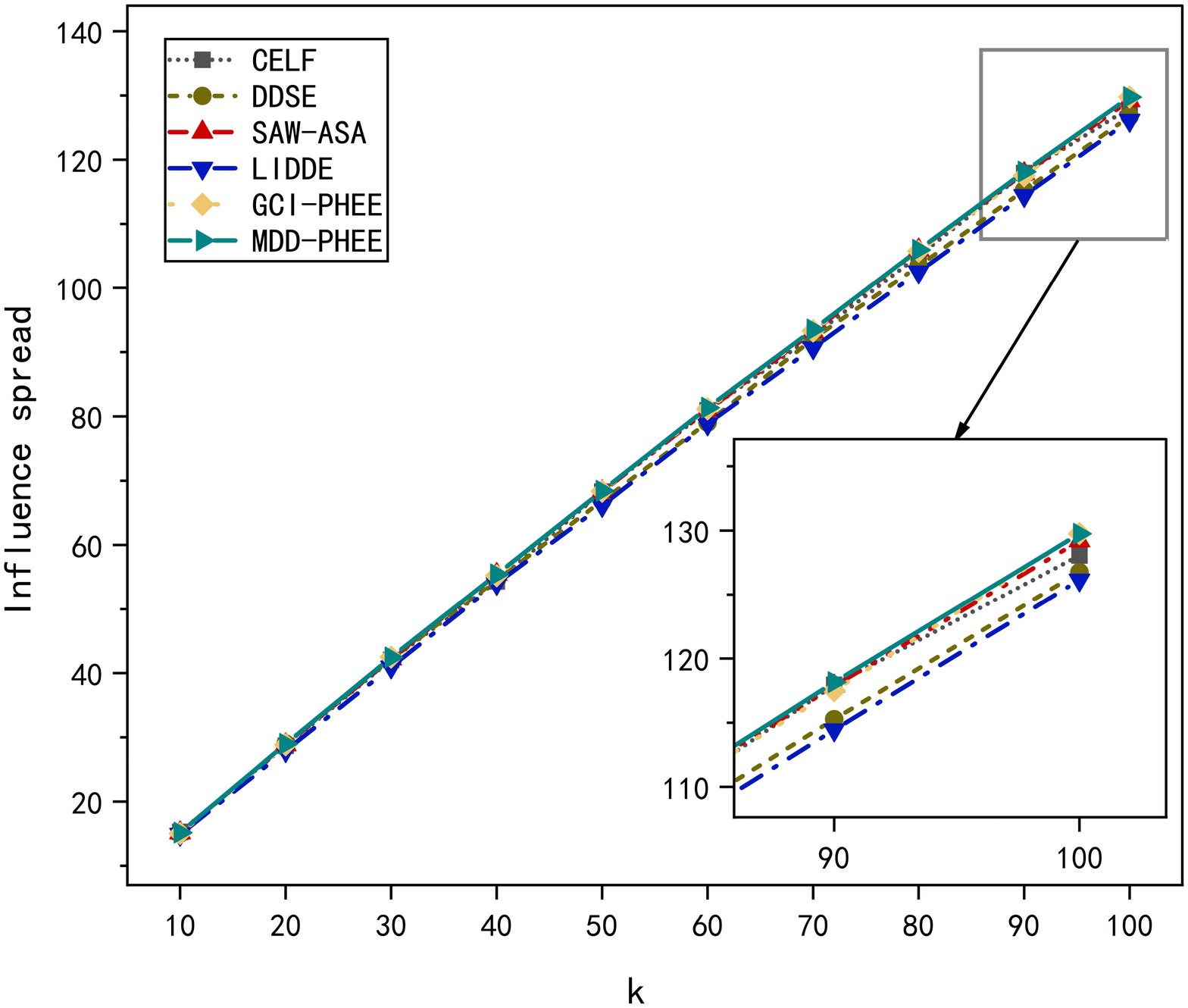}
}
\subfigure[CA-AstroPh]{
\includegraphics[width=7cm]{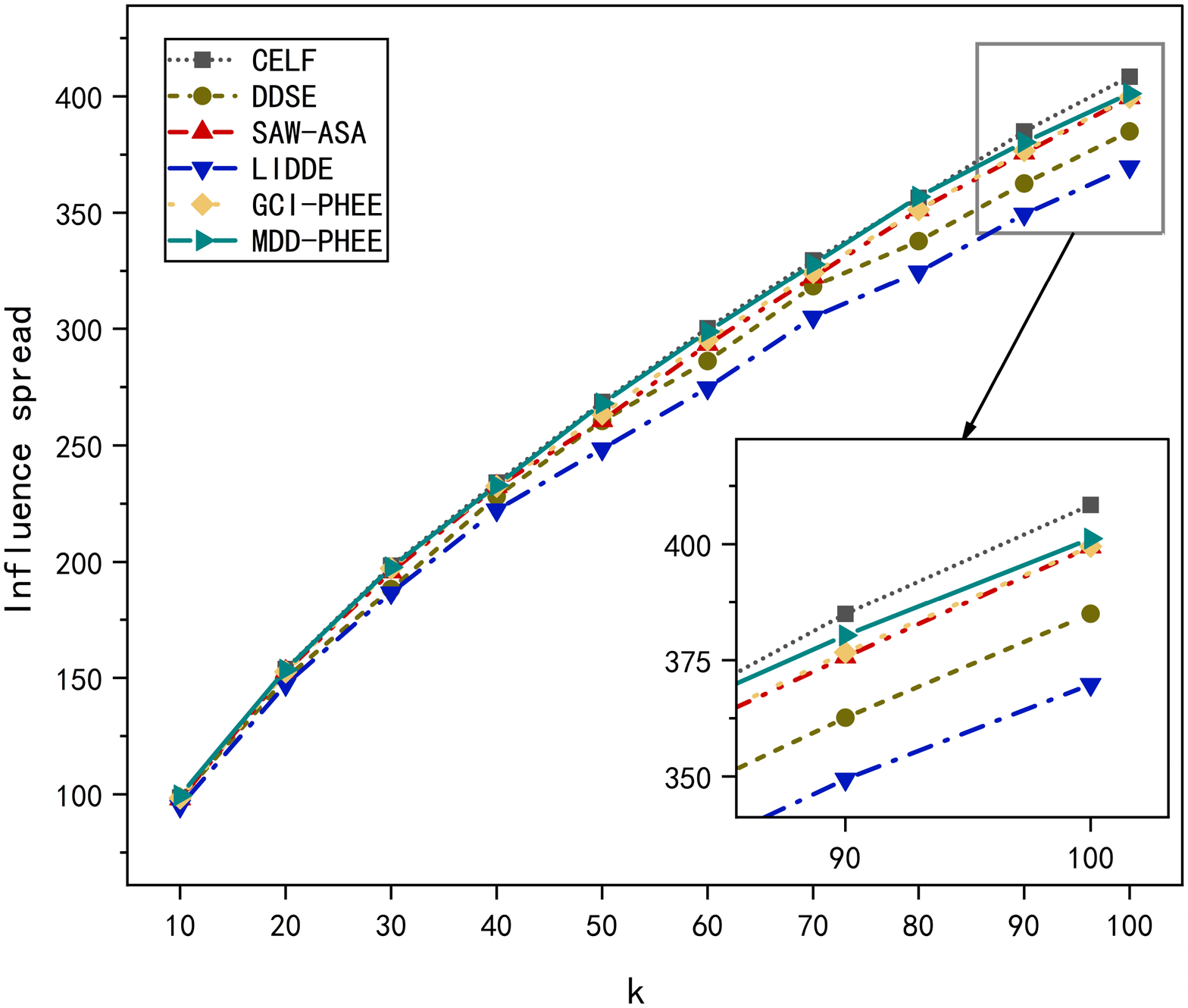}
}
\subfigure[Gnutella]{
\includegraphics[width=7cm]{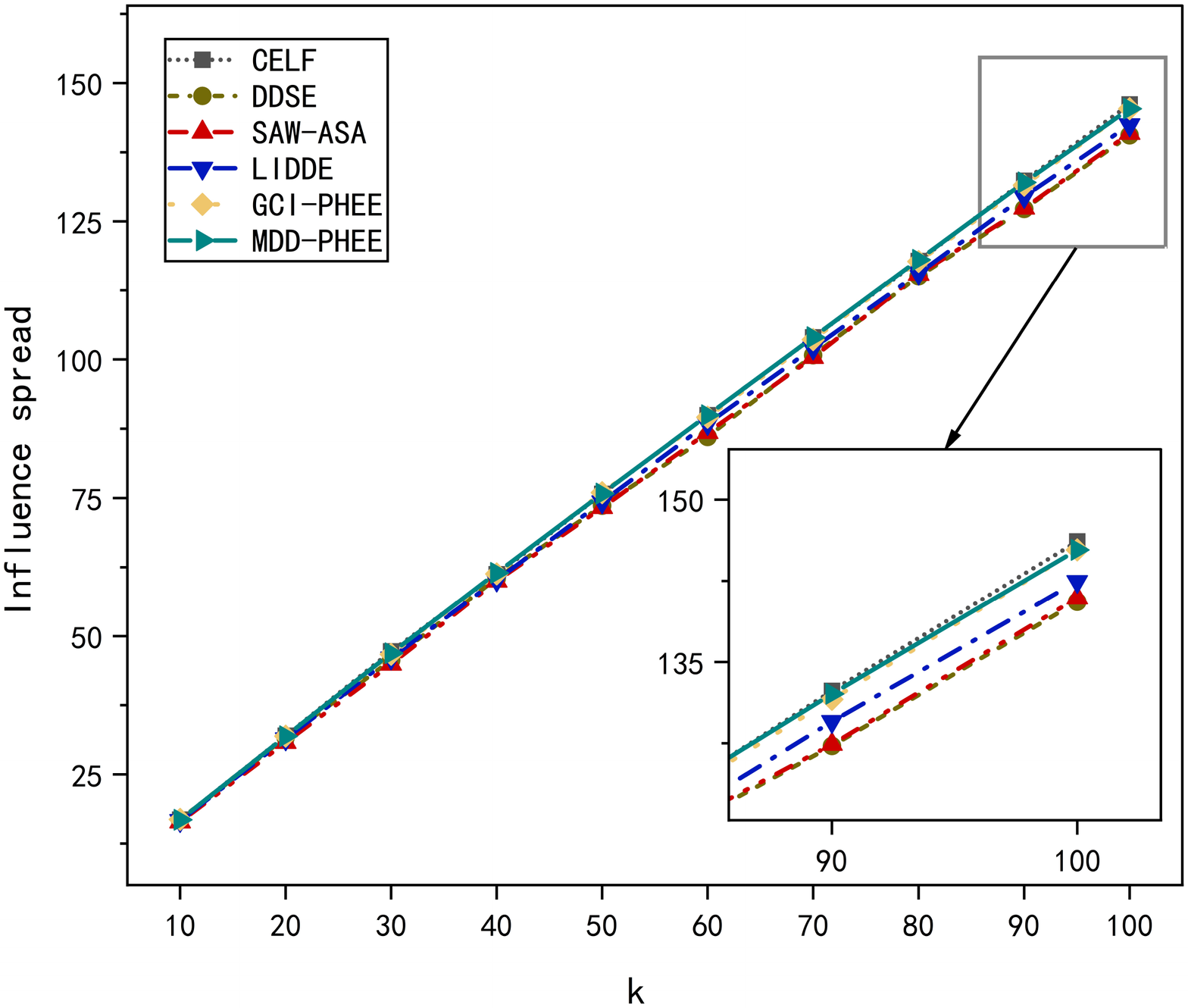}
}
\caption{The comparison on influence spread of different algorithms under IC model.}
\label{figure-exp-1}
\end{center}
\end{figure}

\begin{figure}[H]
\begin{center}
\subfigure[soc-Epinions1]{
\includegraphics[width=7cm]{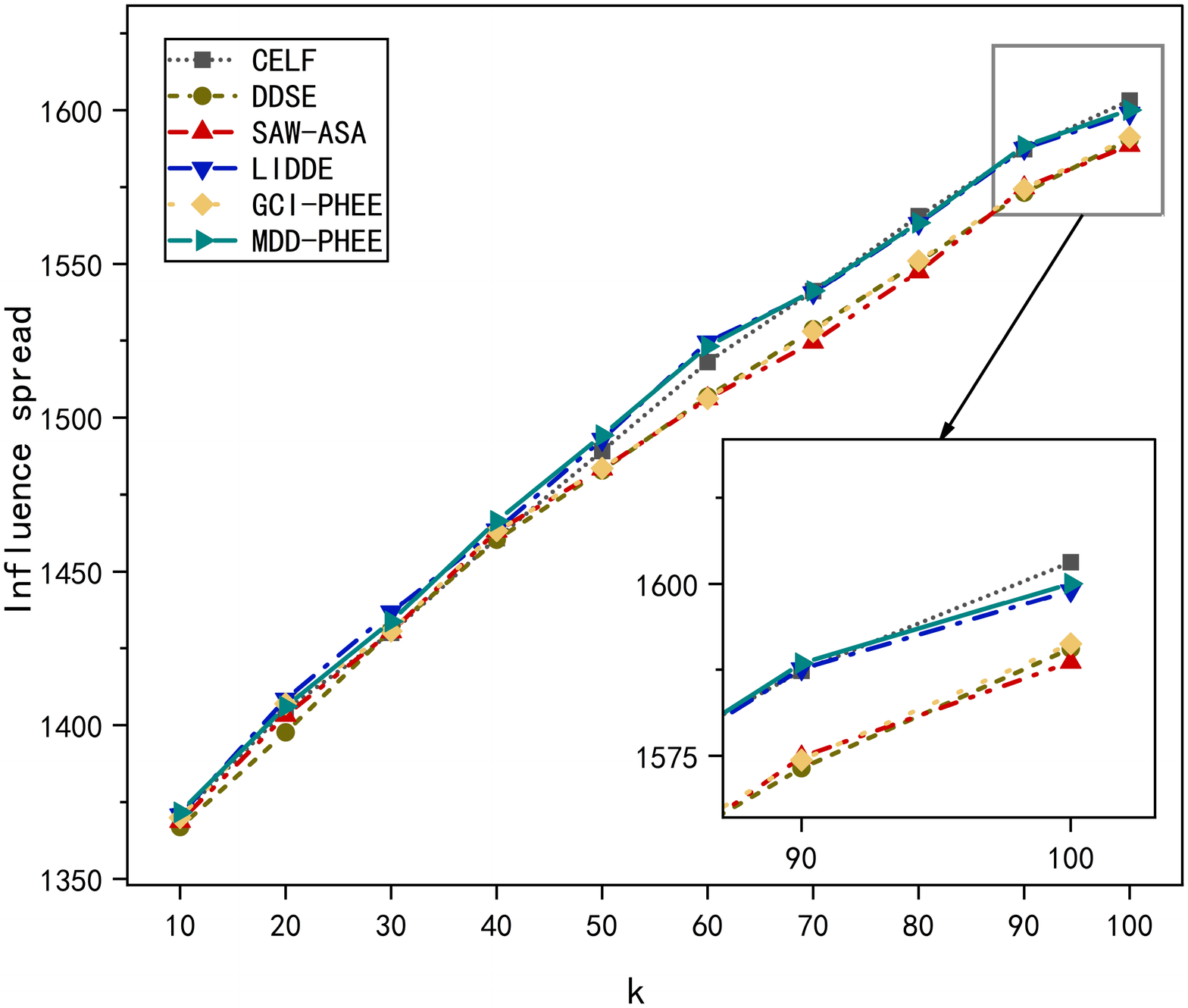}
}
\subfigure[soc-Epinions2]{
\includegraphics[width=7cm]{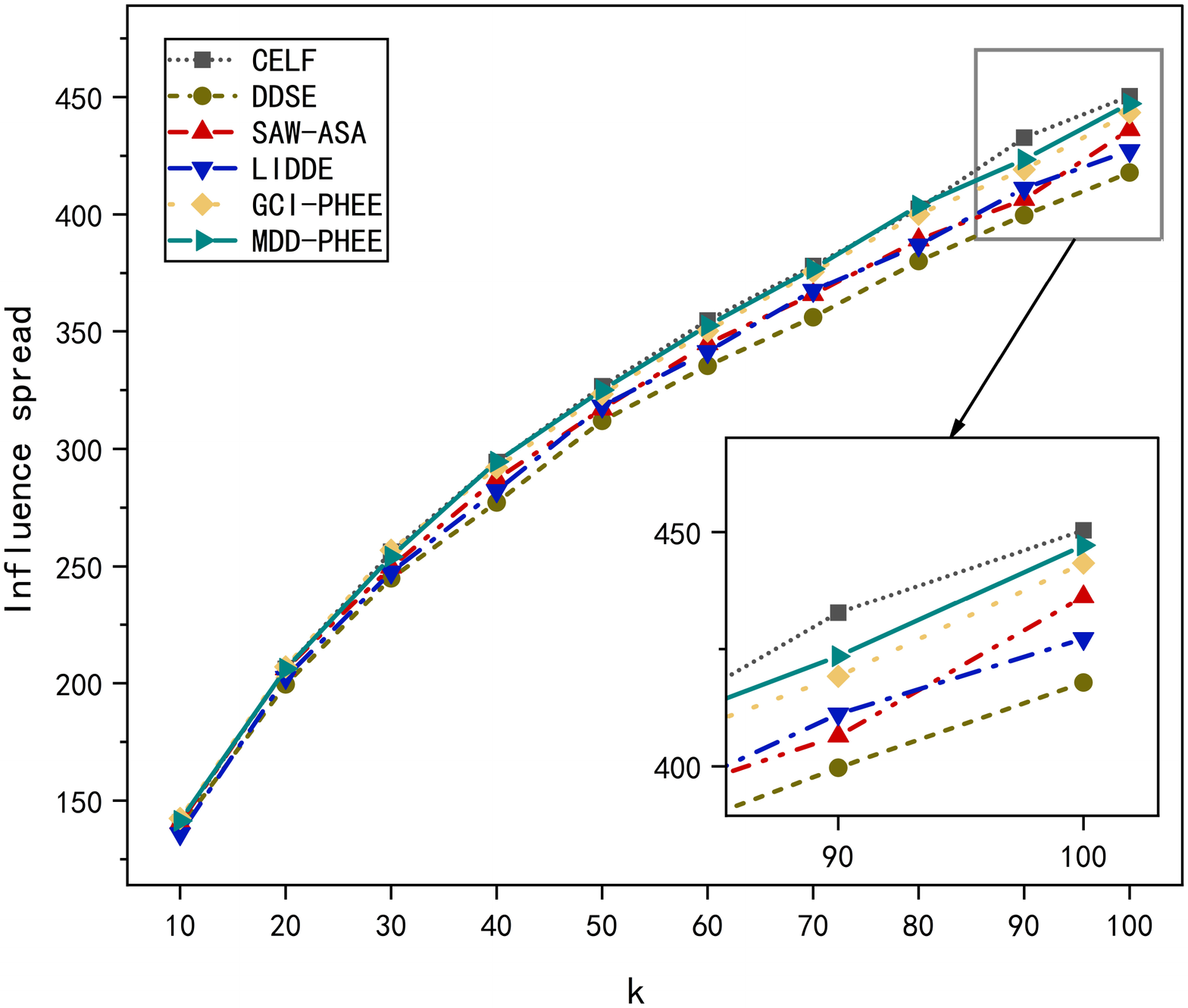}
}
\subfigure[Slashdot]{
\includegraphics[width=7cm]{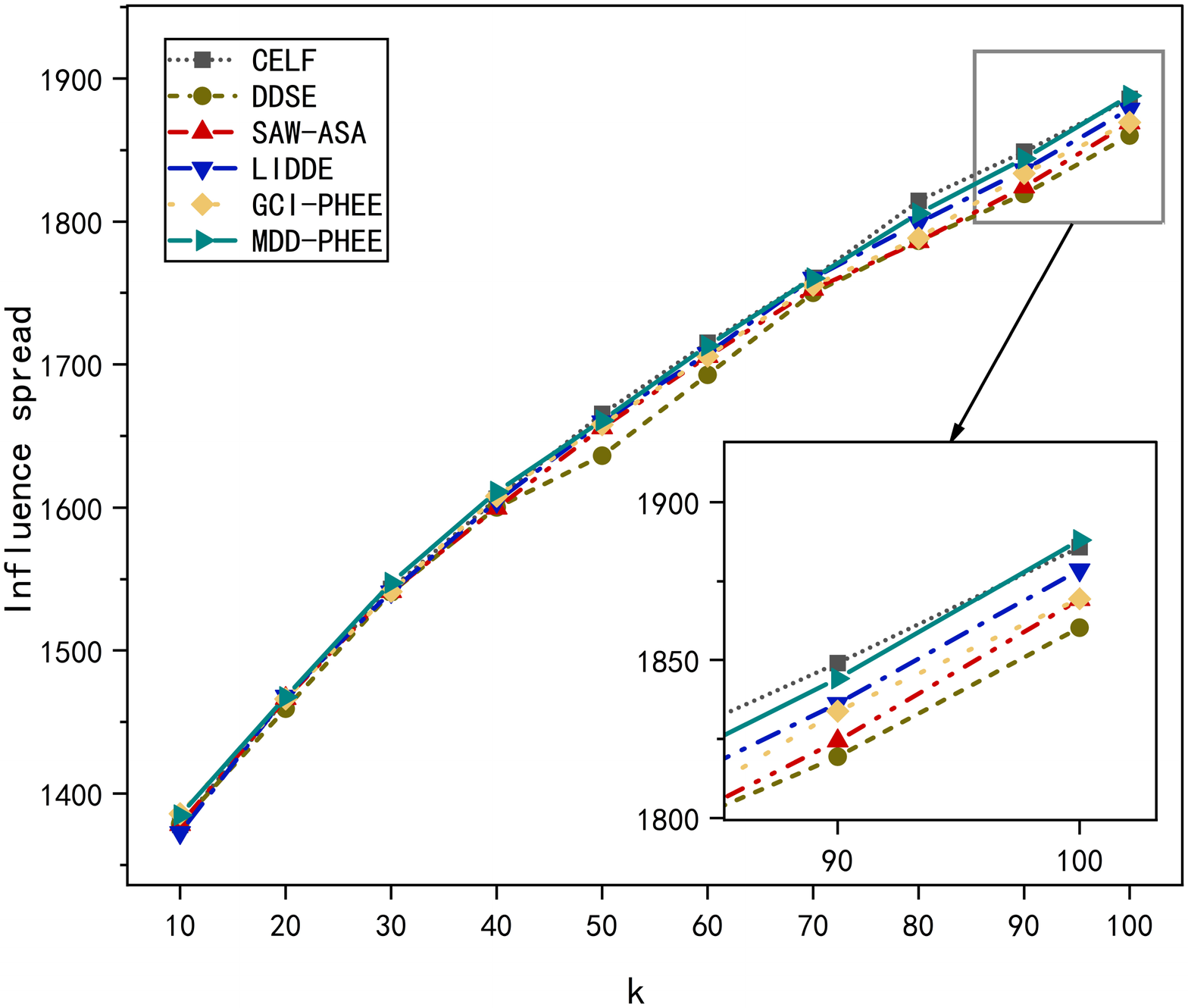}
}
\subfigure[Email-eu]{
\includegraphics[width=7cm]{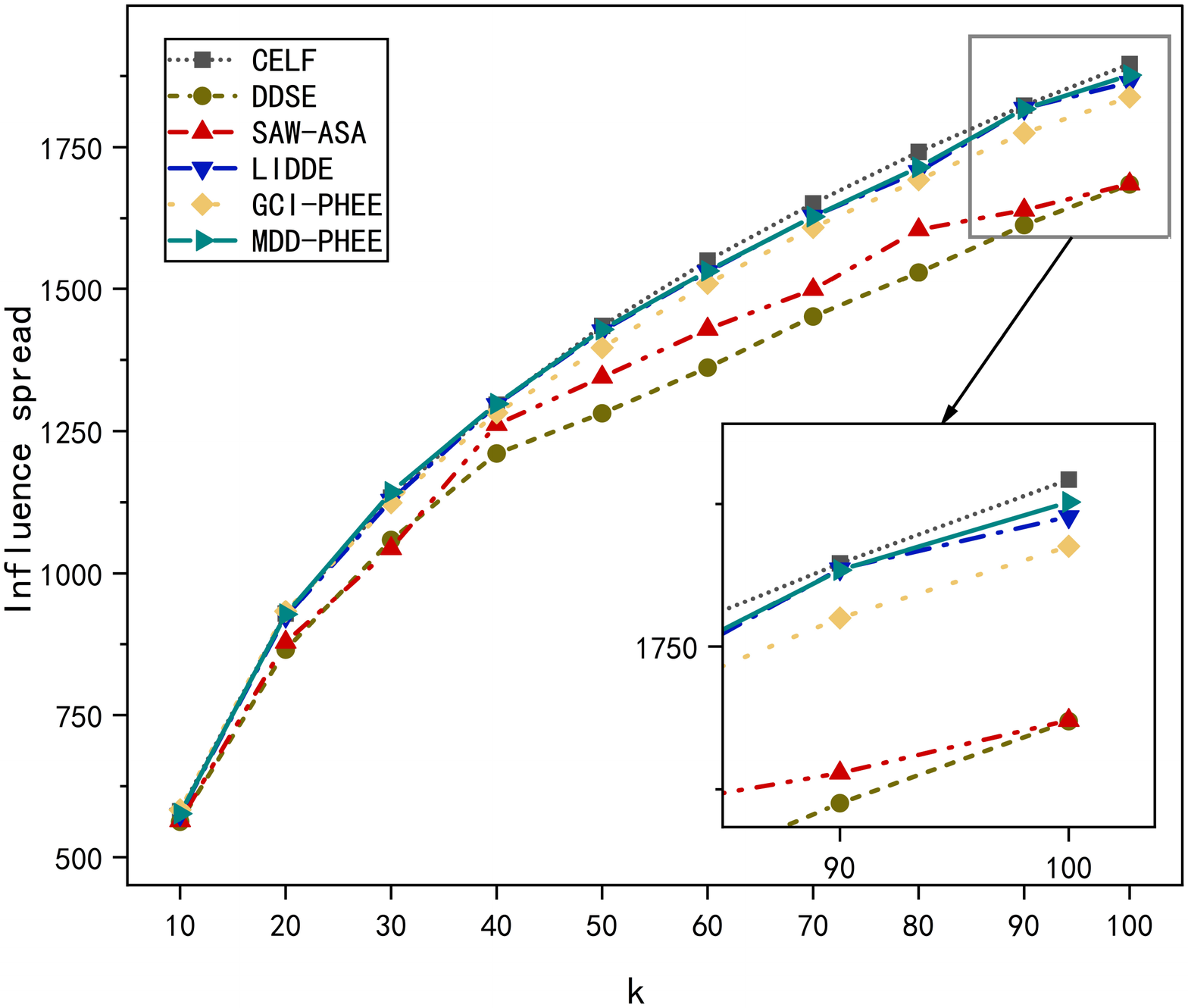}
}
\caption{The comparison on influence spread of different algorithms under IC model.
}
\label{figure-exp-2}
\end{center}
\end{figure}

Figures \ref{figure-exp-1} and \ref{figure-exp-2}  describe the experimental results of the six algorithms on the 10 networks,
 where the horizontal coordinate represents the size $k$ of the seed set,  and the vertical coordinate is the corresponding influence spread. From the results, we see that MDD-PHEE outperforms most   algorithms on all the datasets, except for CELF on very few networks; in particular, the advantage becomes clear with the sizes of networks and   seed sets increasing.

Since Netscience and Email-un are small graphs with a very low average degree, all algorithms have a similar influence spread,  but GCI-PHEE, SAW-ASA, and especially  MDD-PHEE   gradually show their advantages in terms of influence spread with the increase of $k$, and to some extent, it  still can be seen  that MDD-PHEE outperforms other algorithms. We see that MDD-PHEE, GCI-PHEE, and SAW-ASA outperform  the other algorithms on the academic communication networks CA-GrQc, CA-HepTh, and CA-AstroPh, with the exception that CELF provides a slightly better spread  on CA-AstroPh. Moreover, our PHEE framework algorithms outperform SAW-ASA, especially on CA-HepTh and CA-AstroPh. Our PHEE algorithms and CELF perform similarly on Gnutella, outperforming all others. SAW-ASA does not perform as well as before on this network,
while LIDDE can achieve  relatively good performance.
MDD-PHEE and CELF can achieve a better influence spread than other algorithms on the two Epinions networks and the Slashdot network, while LIDDE produces competitive results  only  on Epinions1, and GCI-PHEE performs well only on Epinions2.
Finally, CELF outperforms the others on the large network Email-eu, followed by MDD-PHEE, LIDDE, and MDD-PHEE, while  GCI-PHEE has an influence spread close to theirs.

Overall,  DDSE does not perform  so well on all datasets, possibly because it only uses a single vertex influence estimation strategy.
  SAW-ASA performs well for small networks, but starts to deteriorate as the size increases. This is perhaps due to its single strategy of optimal solution search, i.e., simulated annealing, which limits the diversity of its solutions on larger networks.
   LIDDE   shows great fluctuations of influence spread on  some networks.  For example, it performs poorly on the three academic communication networks, but is competitive on large networks such as Epinions1 and Email-eu.
In contrast, our PHEE algorithms
are relatively stable, since PHEE adopts   different strategies for the  solution space and convergence to the optimal solution. In particular, MDD-PHEE outperforms the other algorithms on almost all of the  networks. This indicates that MDD-based vertex influence estimation is   suitable for PHEE, since changes in the structures and sizes of networks do not result in large fluctuations in the  algorithm's performance. GCI-PHEE is only fit for certain network structures, such as the three academic exchange networks and the Netscience network. Finally, CELF is still  competitive on most networks, due to its greedy properties.

\begin{figure}[H]
\begin{center}
\subfigure[Netscience]{
\includegraphics[width=3.5cm]{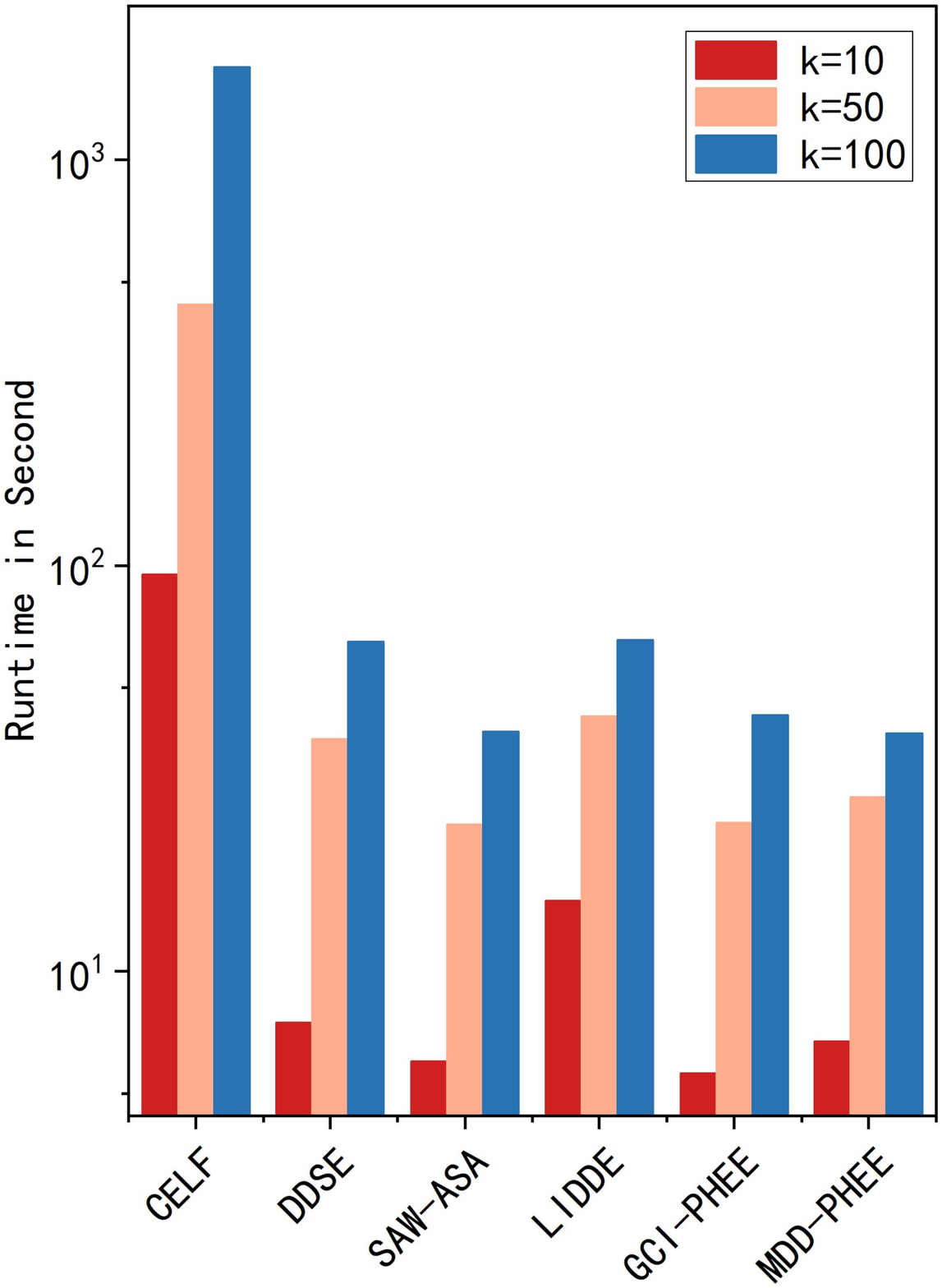}
}
\subfigure[Email-un]{
\includegraphics[width=3.5cm]{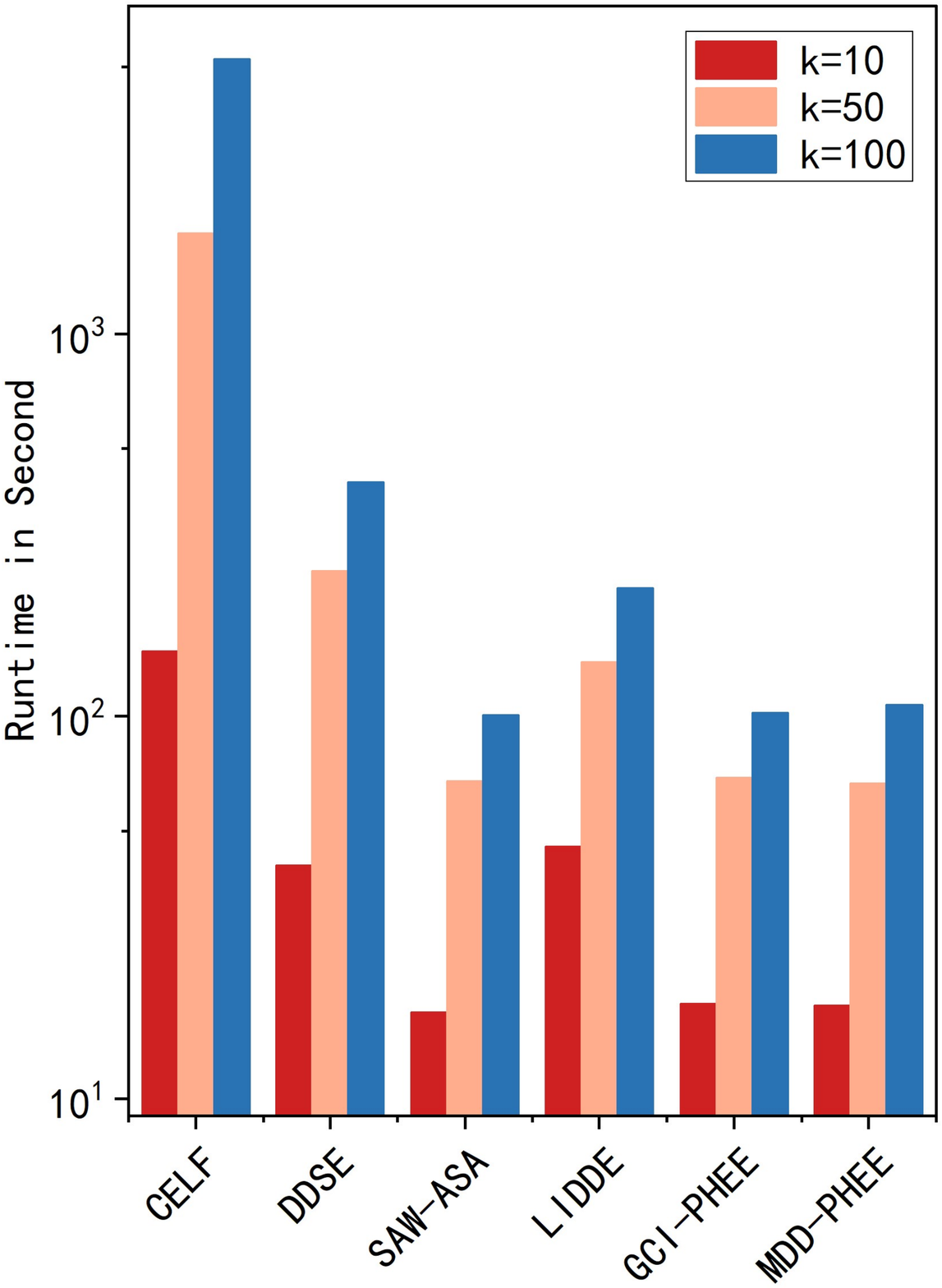}
}
\subfigure[CA-GrQc]{
\includegraphics[width=3.5cm]{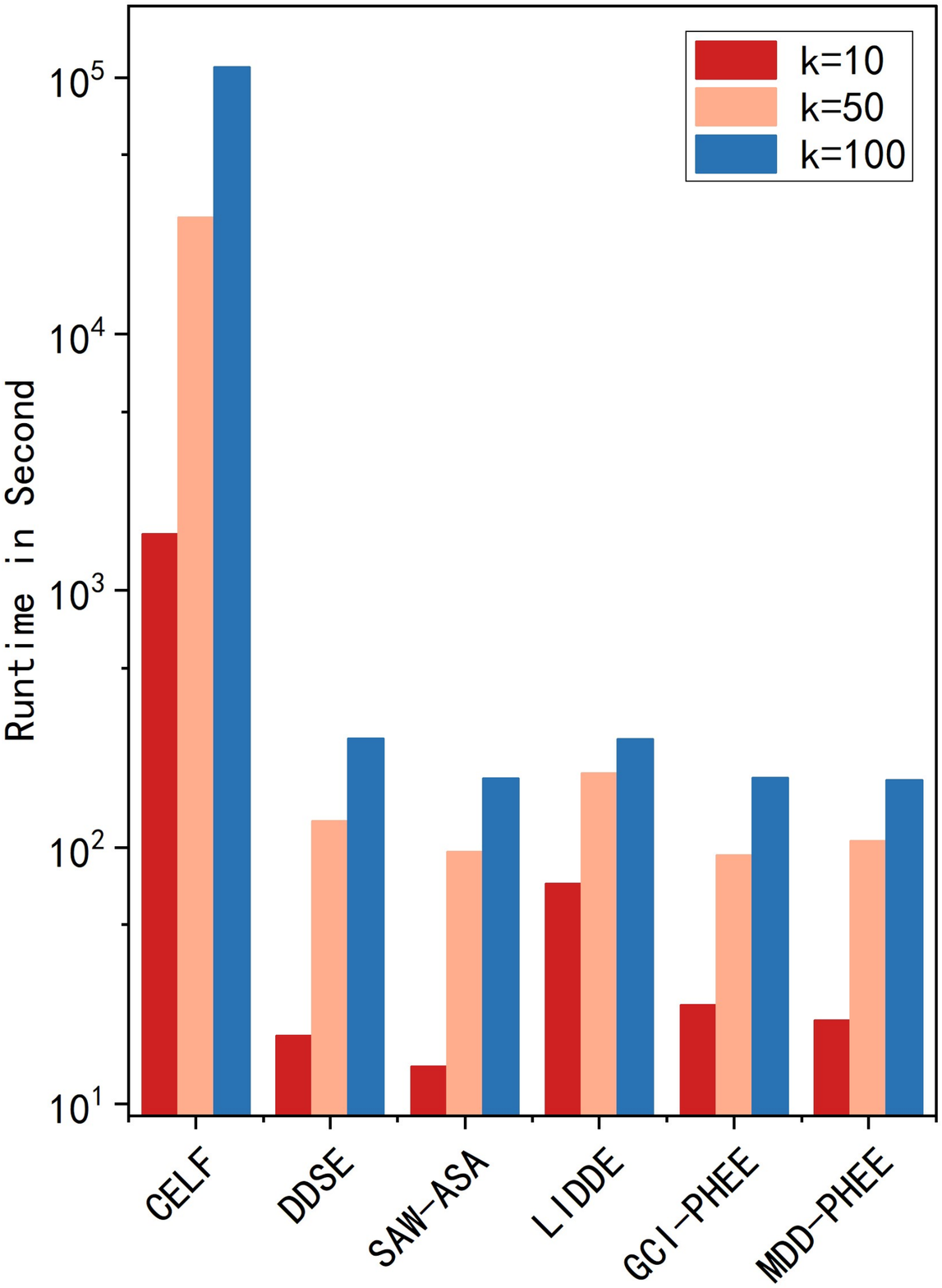}
}
\subfigure[CA-HepTh]{
\includegraphics[width=3.5cm]{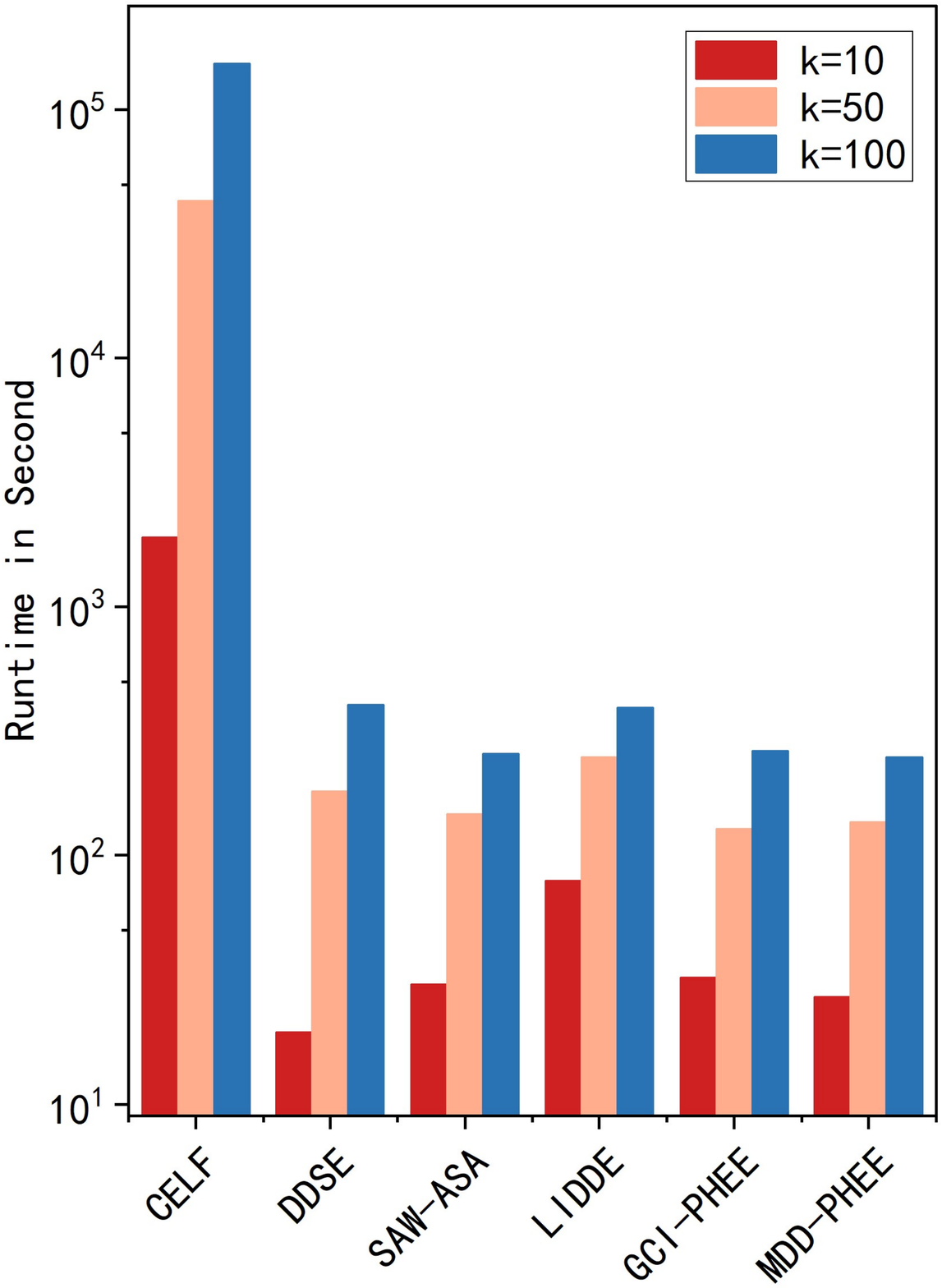}
}\\
\subfigure[CA-AstroPh]{
\includegraphics[width=3.5cm]{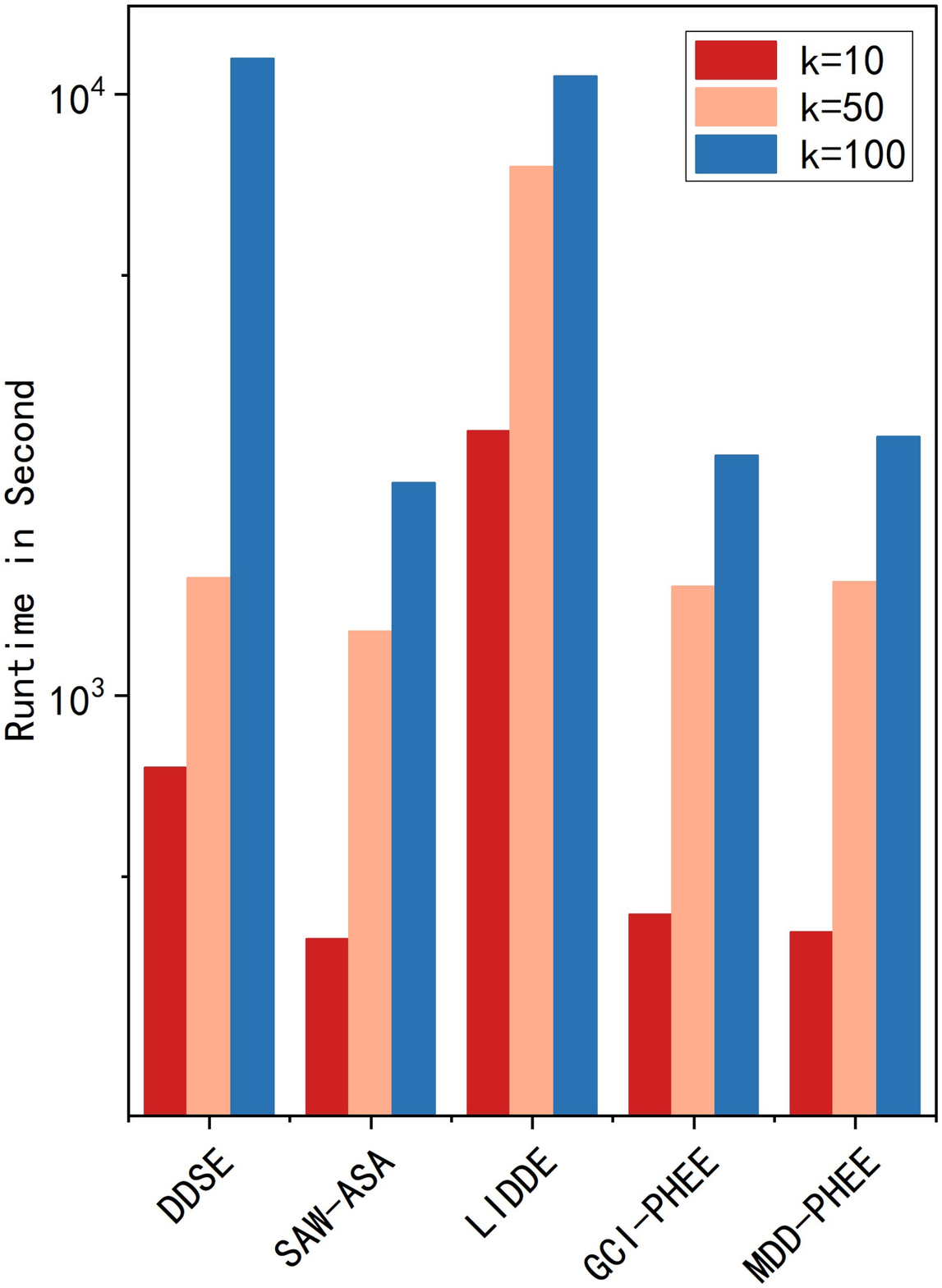}
}
\subfigure[Gnutella]{
\includegraphics[width=3.5cm]{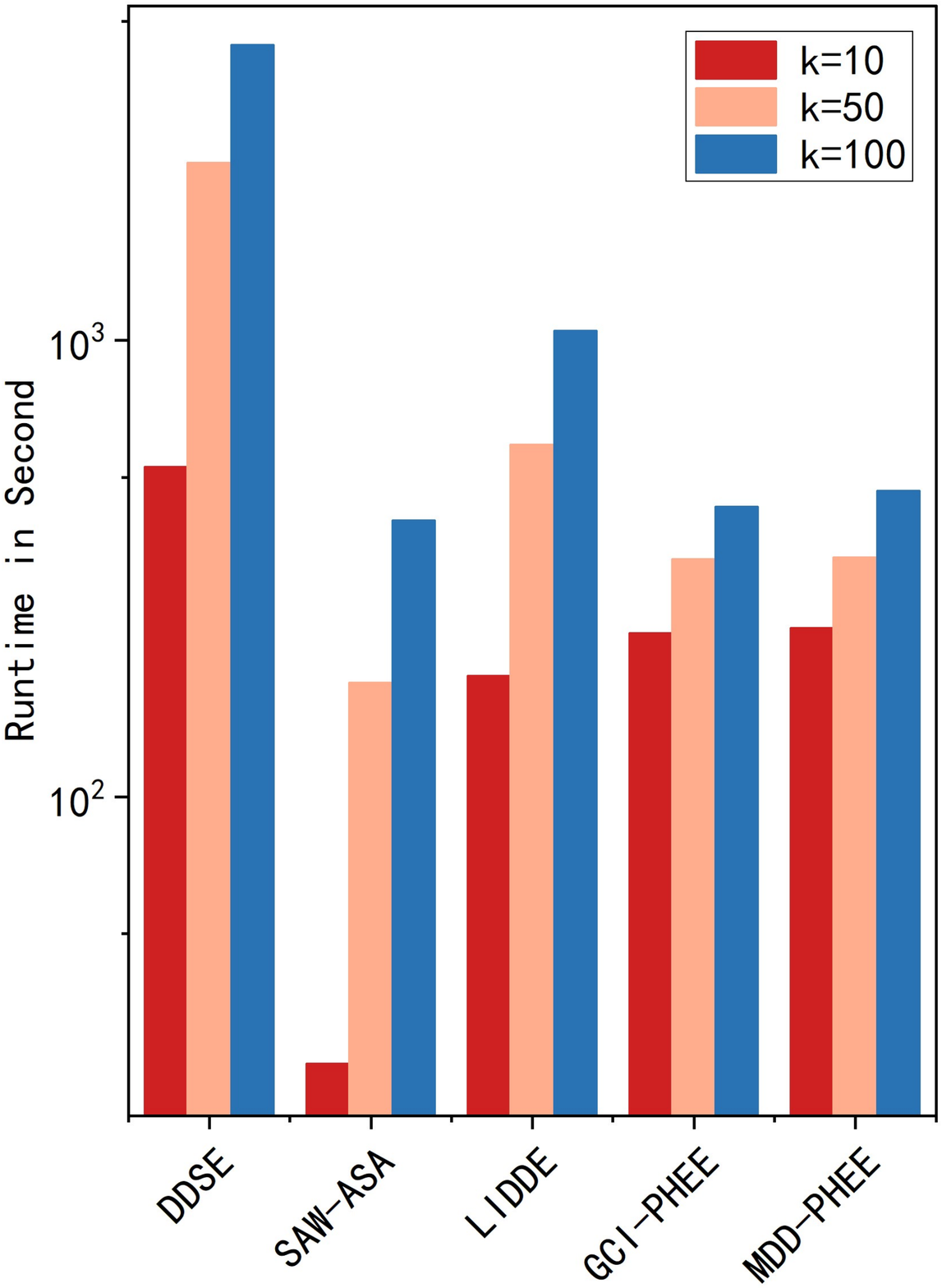}
}
\subfigure[soc-Epinions1]{
\includegraphics[width=3.5cm]{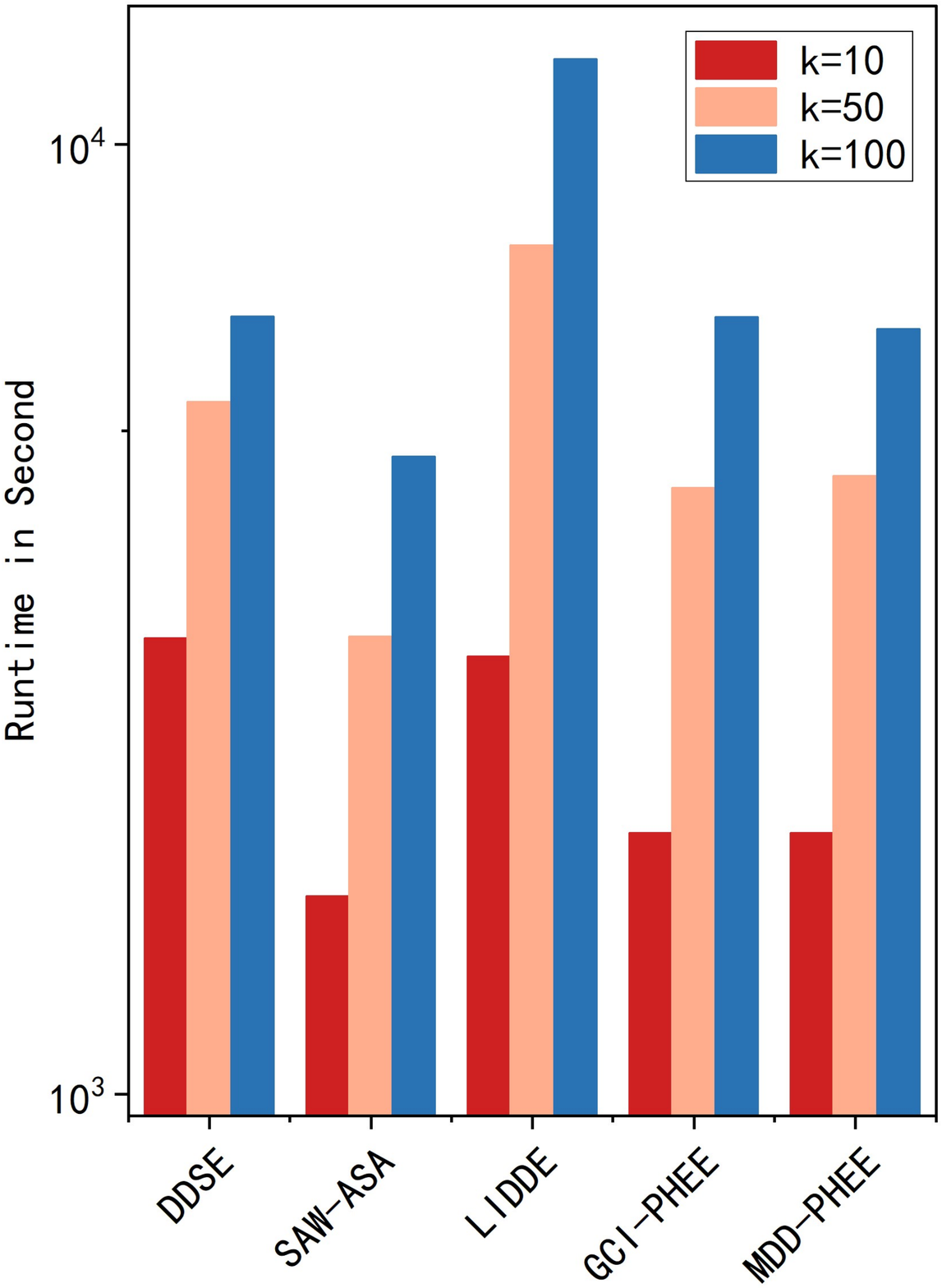}
}
\subfigure[soc-Epinions2]{
\includegraphics[width=3.5cm]{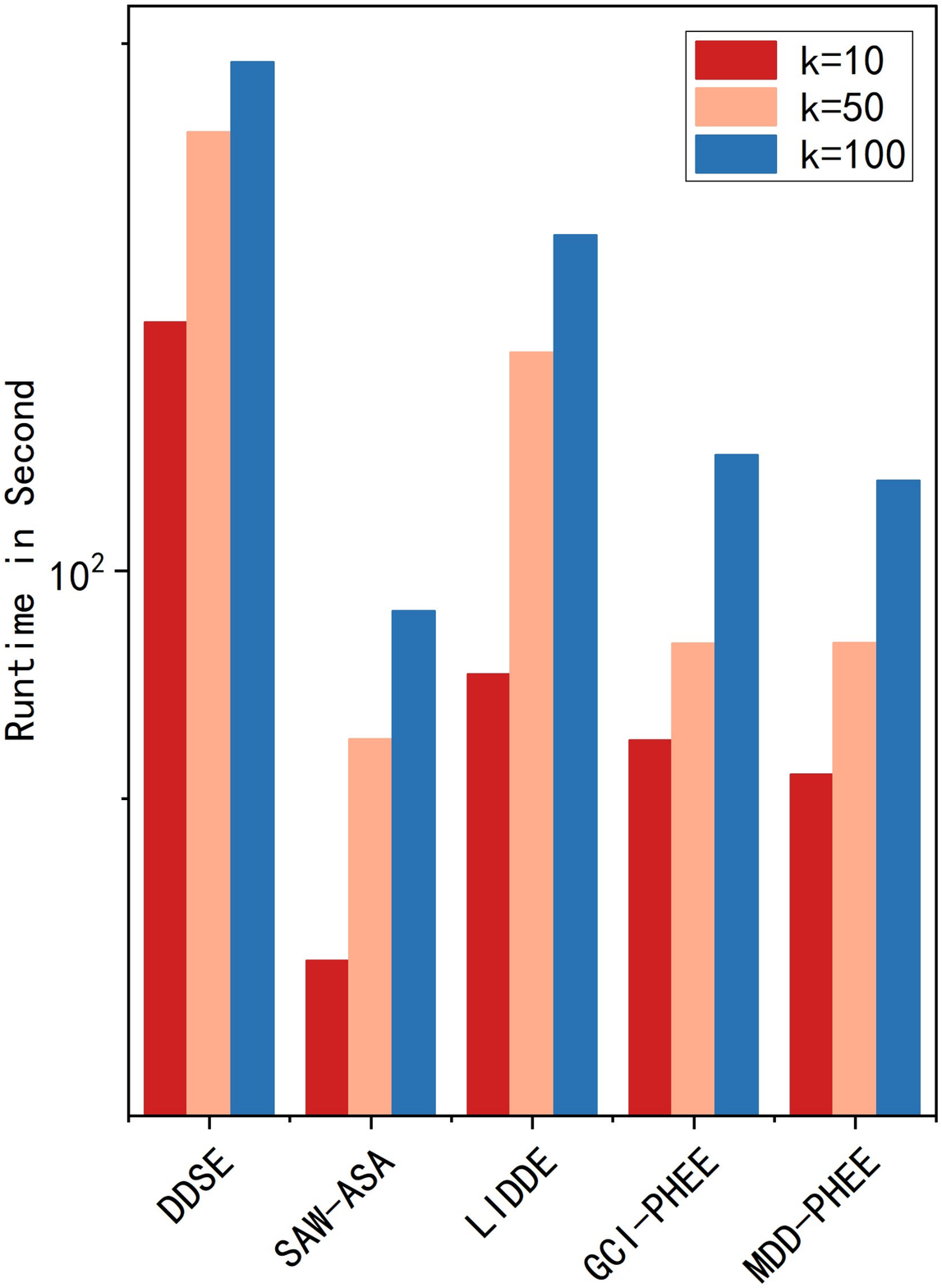}
}\\
\subfigure[Slashdot]{
\includegraphics[width=3.5cm]{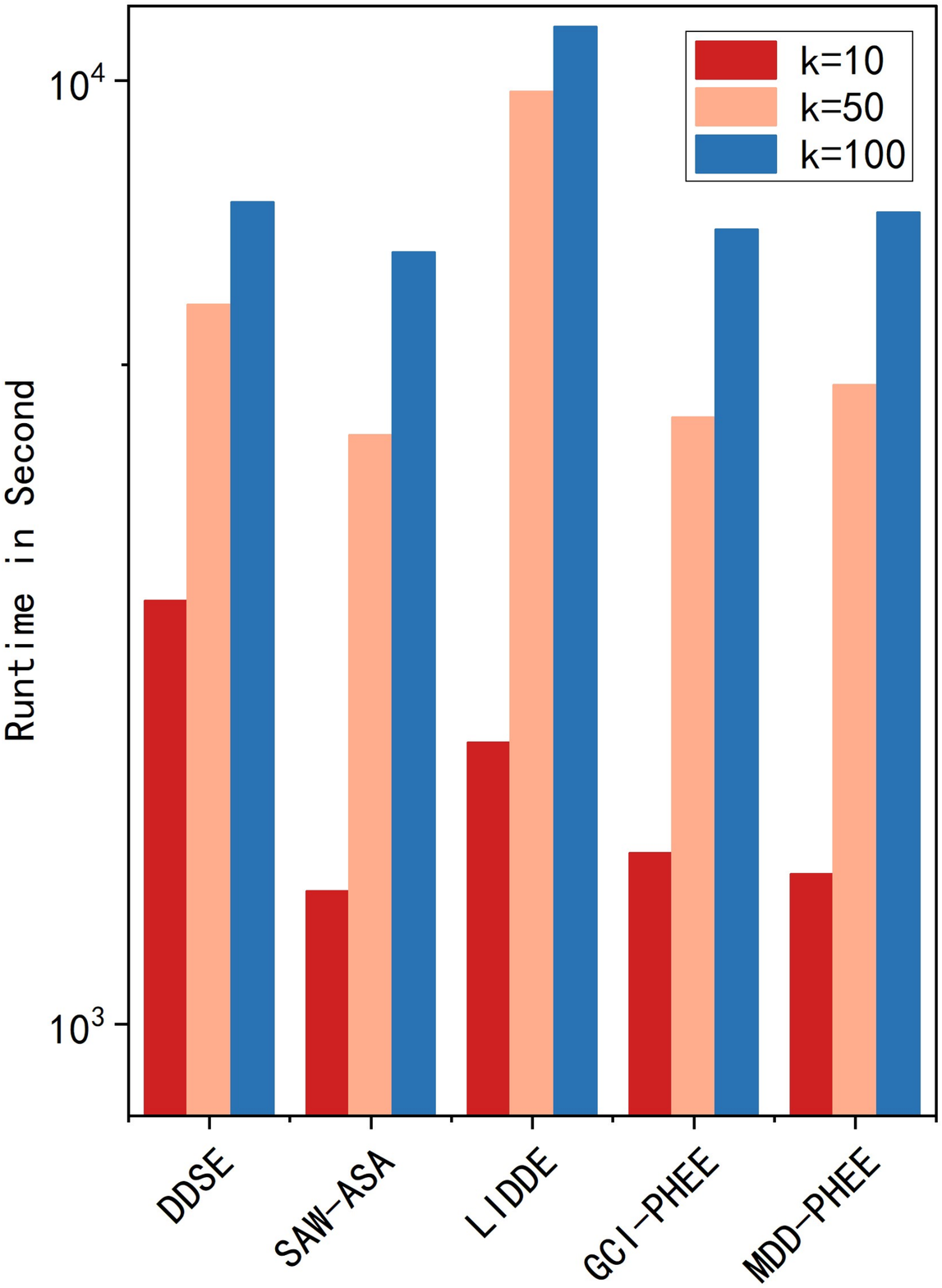}
}
\subfigure[Email-eu]{
\includegraphics[width=3.5cm]{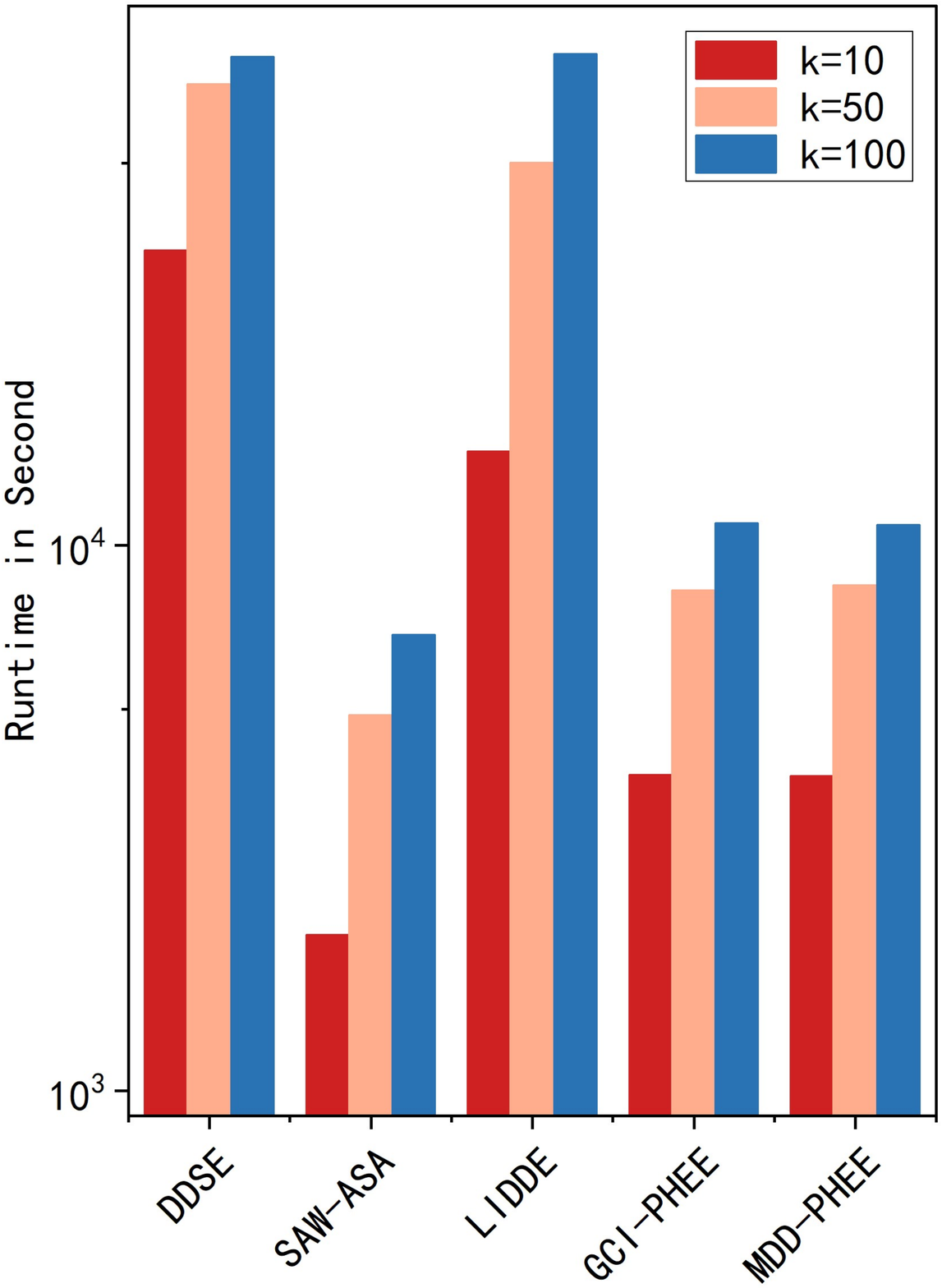}
}
\caption{The comparison on running time of different algorithms under IC model.}
\label{fig-exp-3}
\end{center}
\end{figure}

\subsection{Comparison on running time}

To measure the efficiency of  algorithms, we compared their running times. We recorded  the computational times of each algorithm for calculating the influence spread on the 10 networks with seed sets of size 10, 50, and 100.

The results are  shown in the Figure \ref{fig-exp-3}, where the horizontal axis represents  the algorithms (under the three seed set sizes 10, 50, and 100) for comparison,  and the vertical axis is the running time (seconds). Given that CELF is time-consuming, we only applied it to small networks.

Unsurprisingly, CELF had the worst efficiency on all 10 datasets.  SAW-ASA was the fastest, followed by  GCI-PHEE and MDD-PHEE. Although SAW-ASA was slightly faster than GCI-PHEE and MDD-PHEE, its advantage is not so obvious (these algorithms run in a similar efficiency). The two evolutionary algorithms, DDSE and LIDDE, ran at lower efficiency, especially on large networks, perhaps because of the fitness function used in LIDDE (which was not very efficient) and the restriction of iterations of DDSE (which needs more iterations for solution search). Overall, the PHEE algorithms can produce a high-quality solution in a reasonable time.

\subsection{Self-Comparison Experiment}

\begin{figure}[H]
\begin{center}
\subfigure[Netscience]{
\includegraphics[width=3cm]{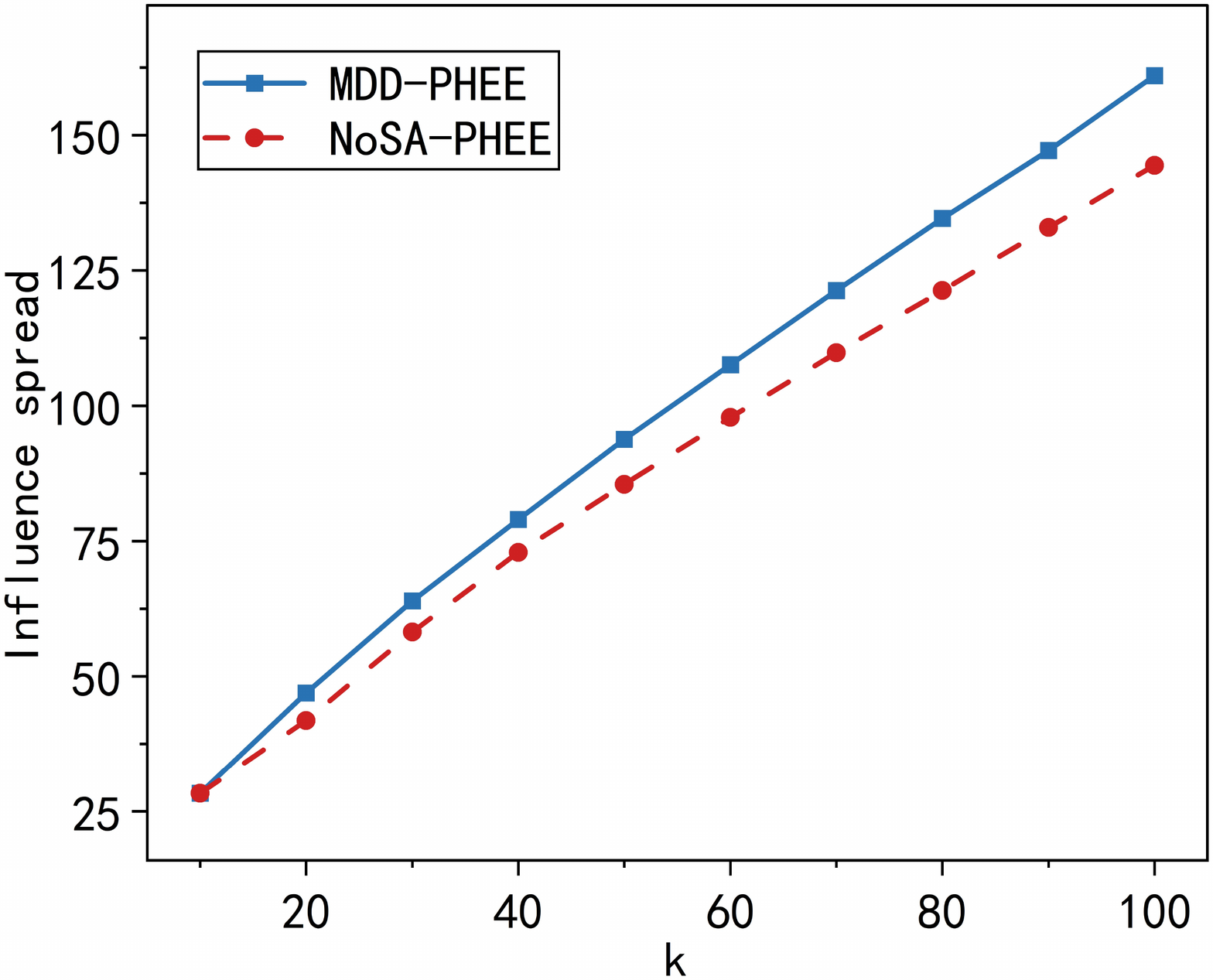}
}
\subfigure[Email-un]{
\includegraphics[width=3cm]{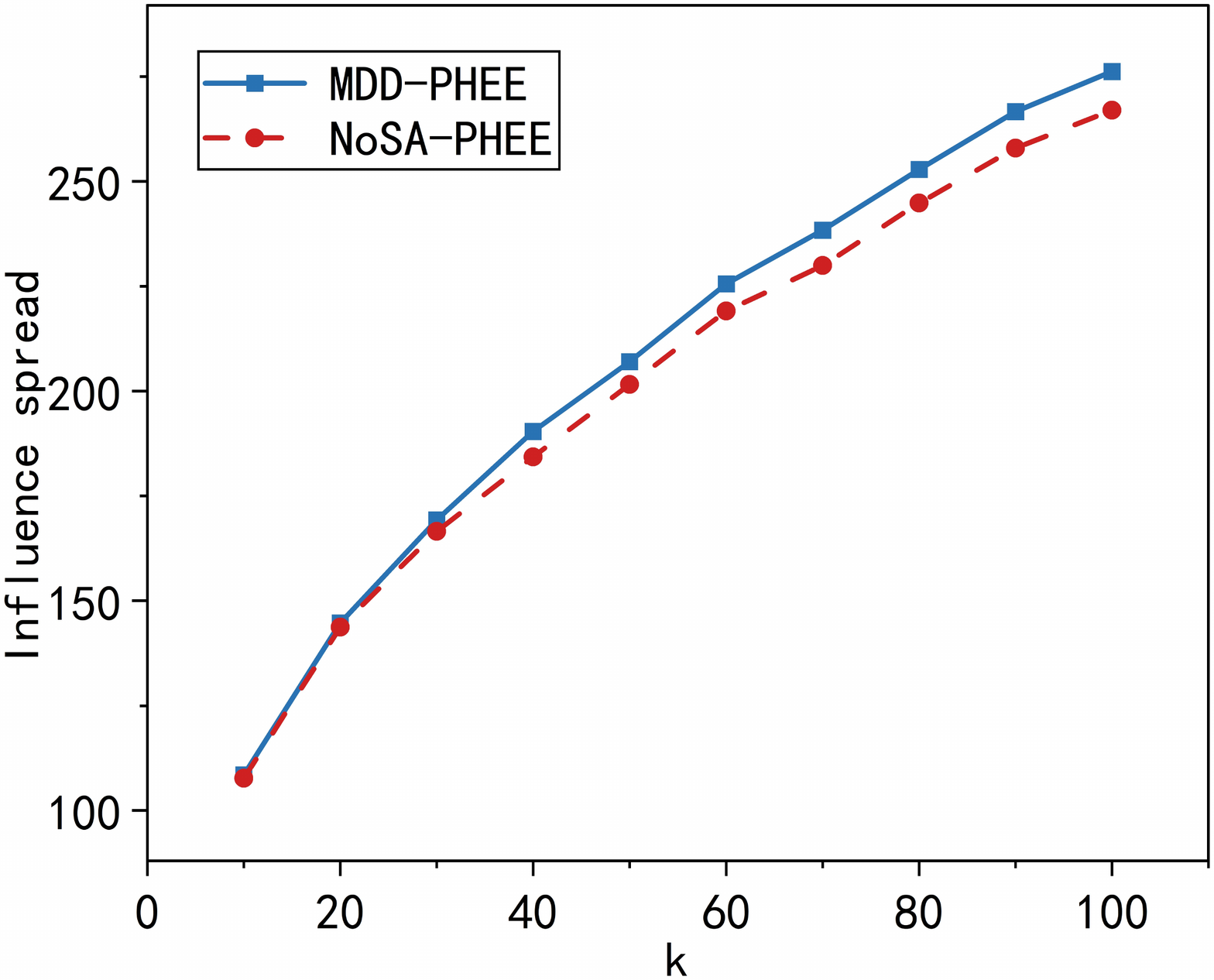}
}
\subfigure[CA-GrQc]{
\includegraphics[width=3cm]{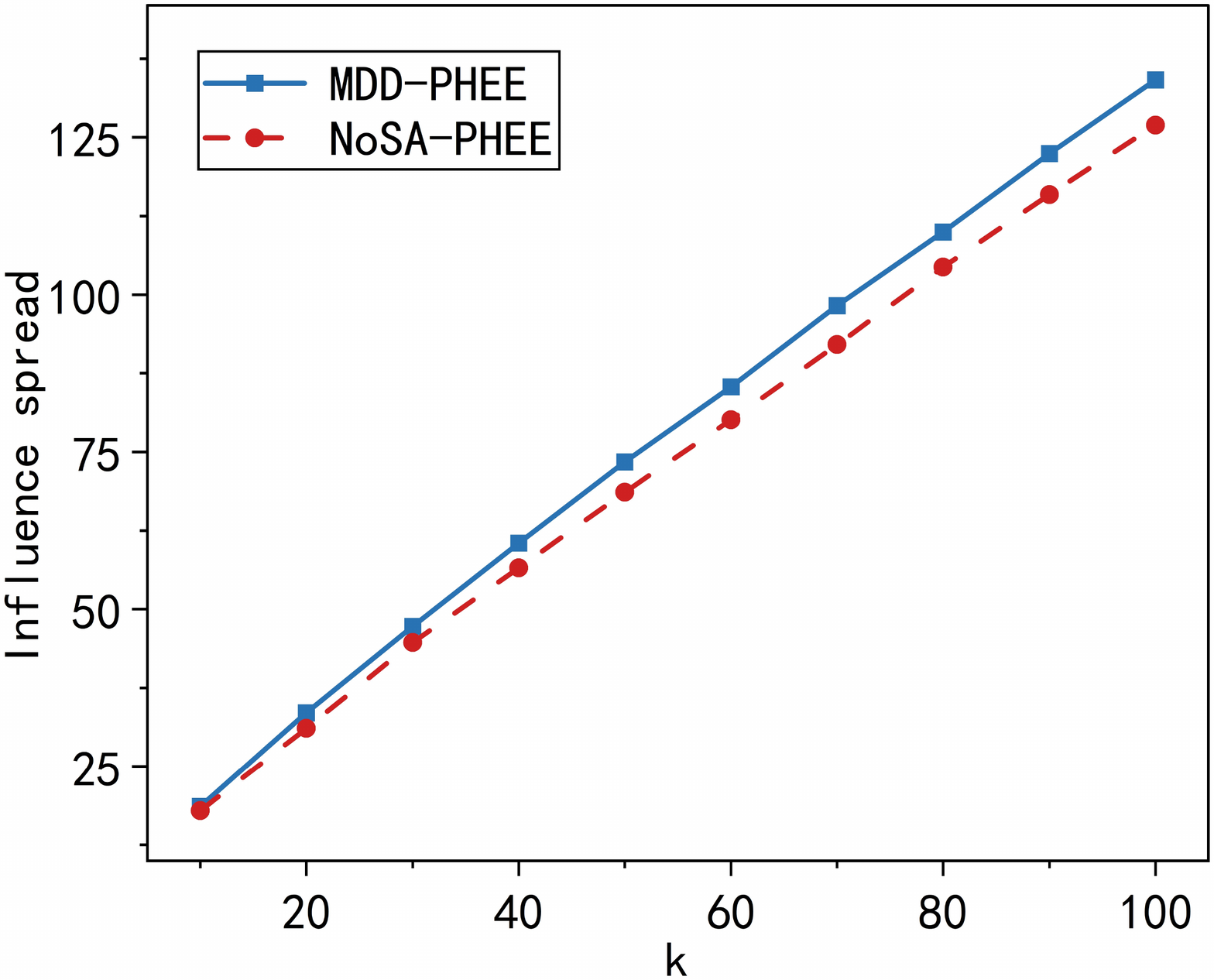}
}
\subfigure[CA-HepTh]{
\includegraphics[width=3cm]{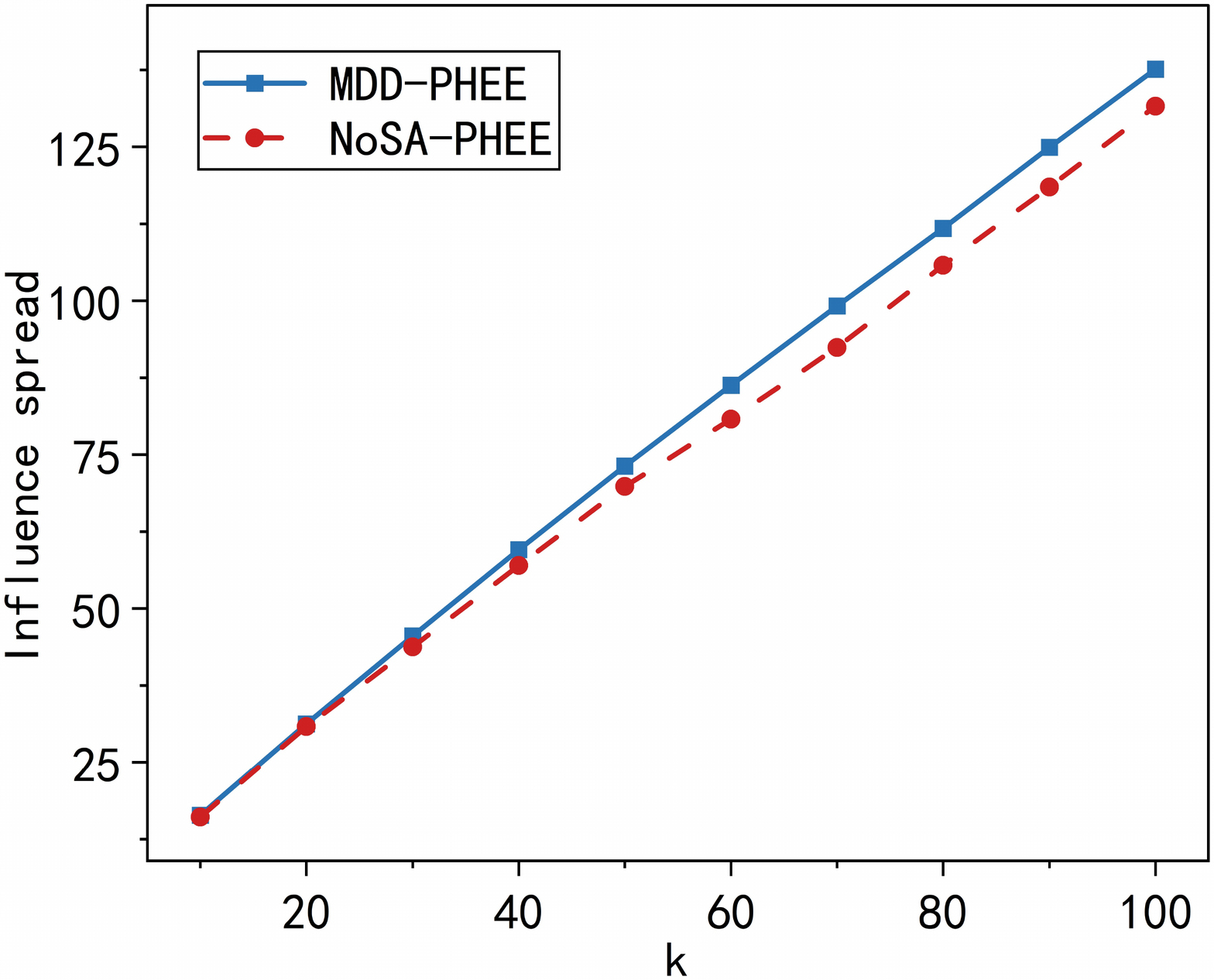}
}
\subfigure[CA-AstroPh]{
\includegraphics[width=3cm]{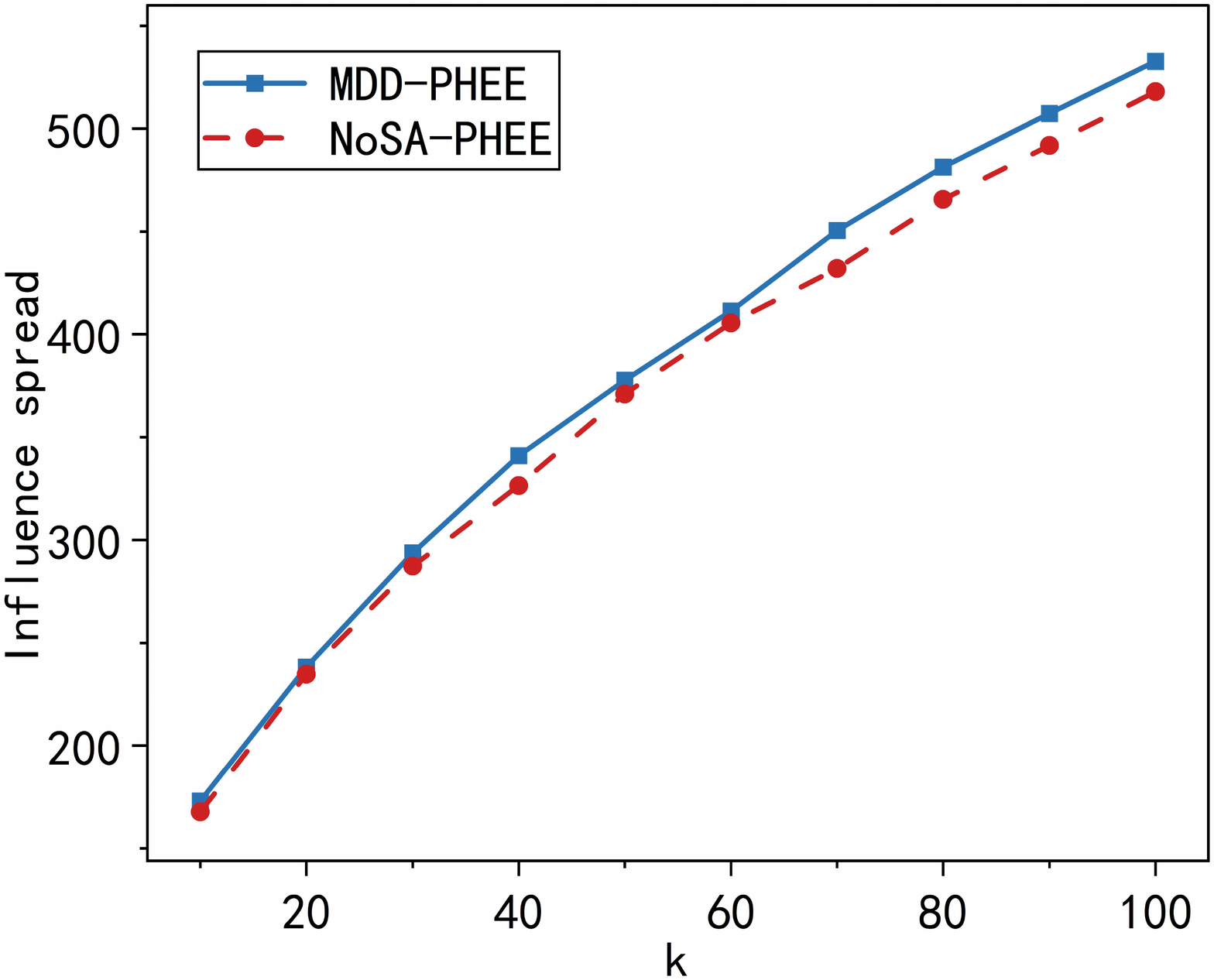}
}\\
\subfigure[Gnutella]{
\includegraphics[width=3cm]{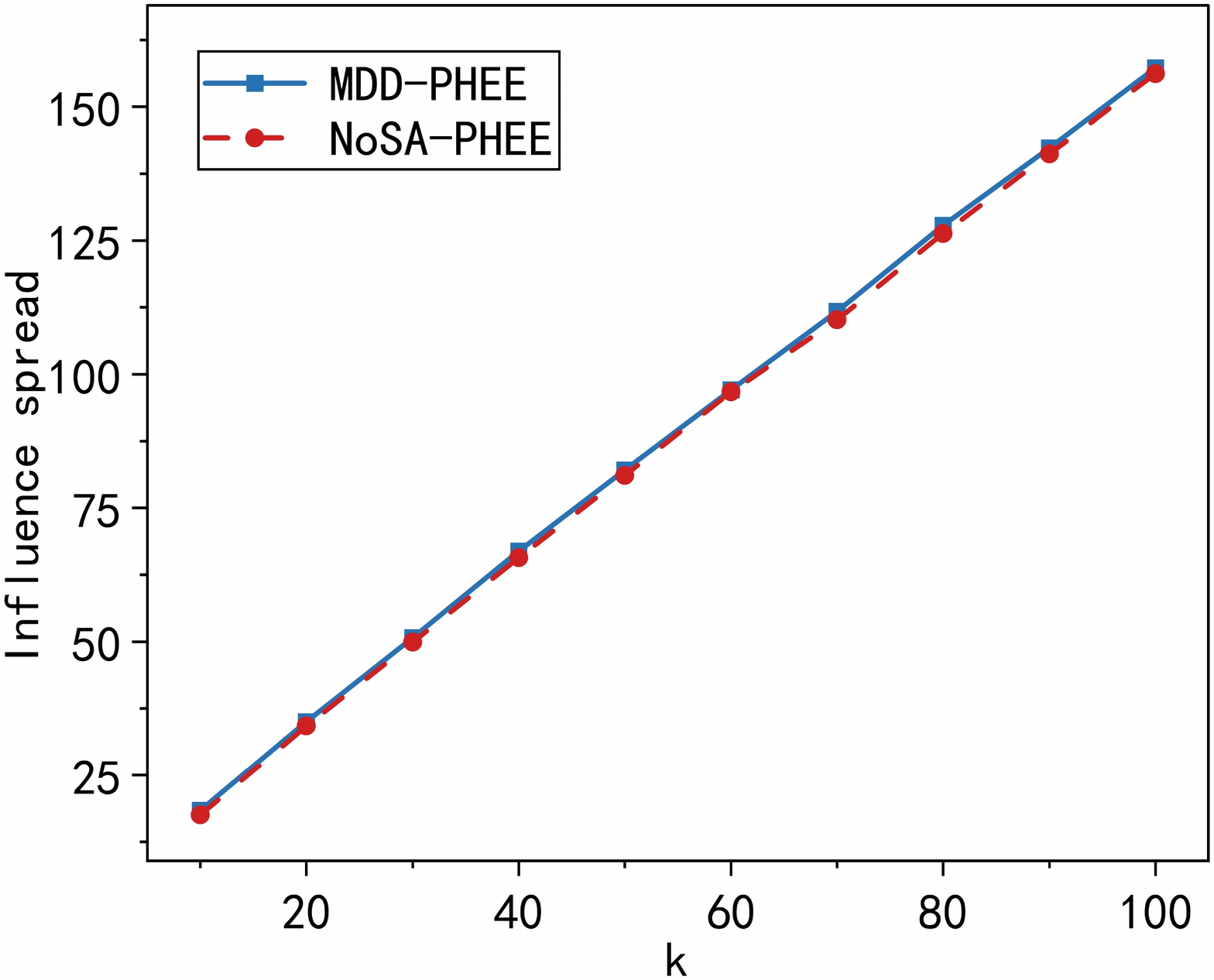}
}
\subfigure[soc-Epinions1]{
\includegraphics[width=3cm]{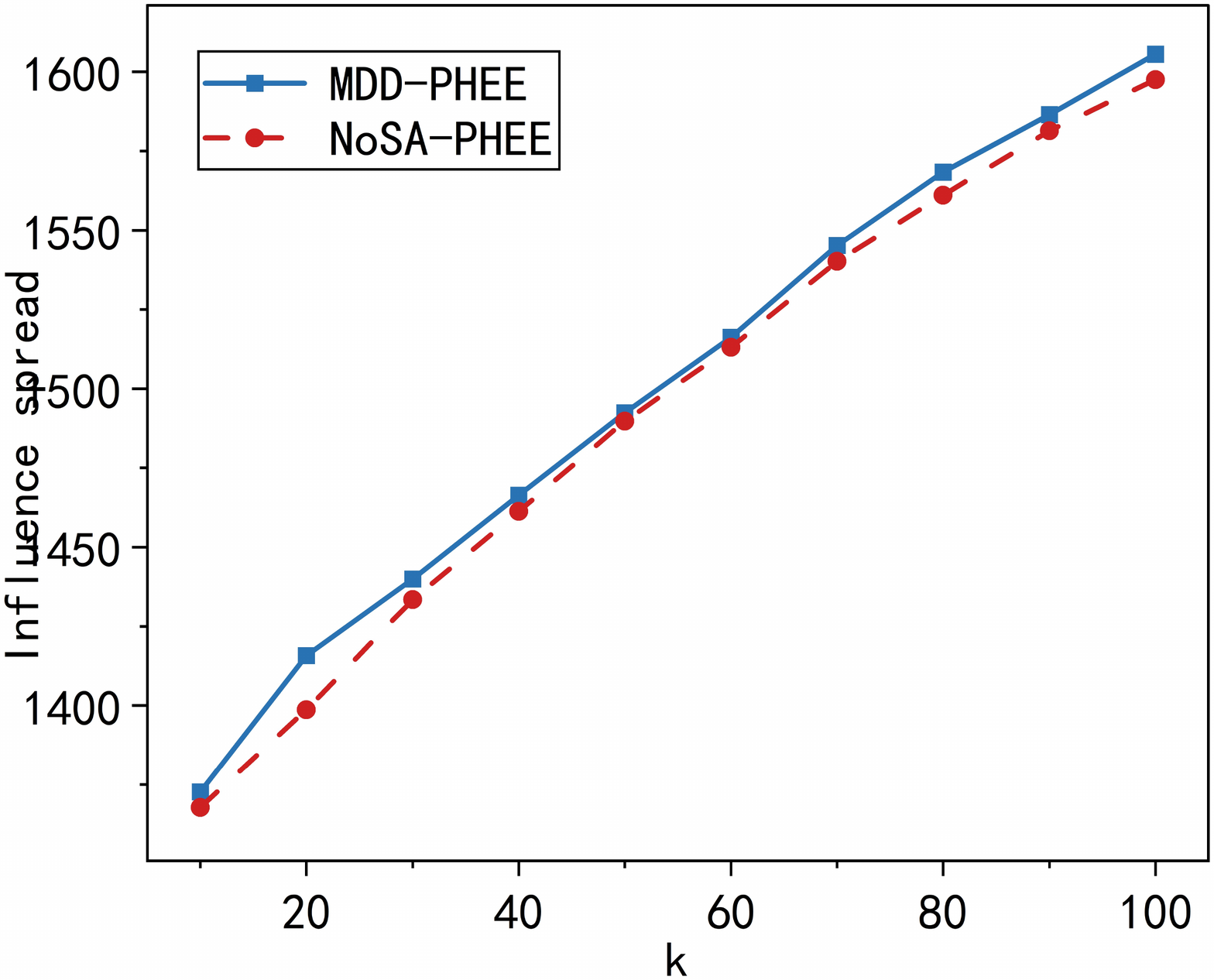}
}
\subfigure[soc-Epinions2]{
\includegraphics[width=3cm]{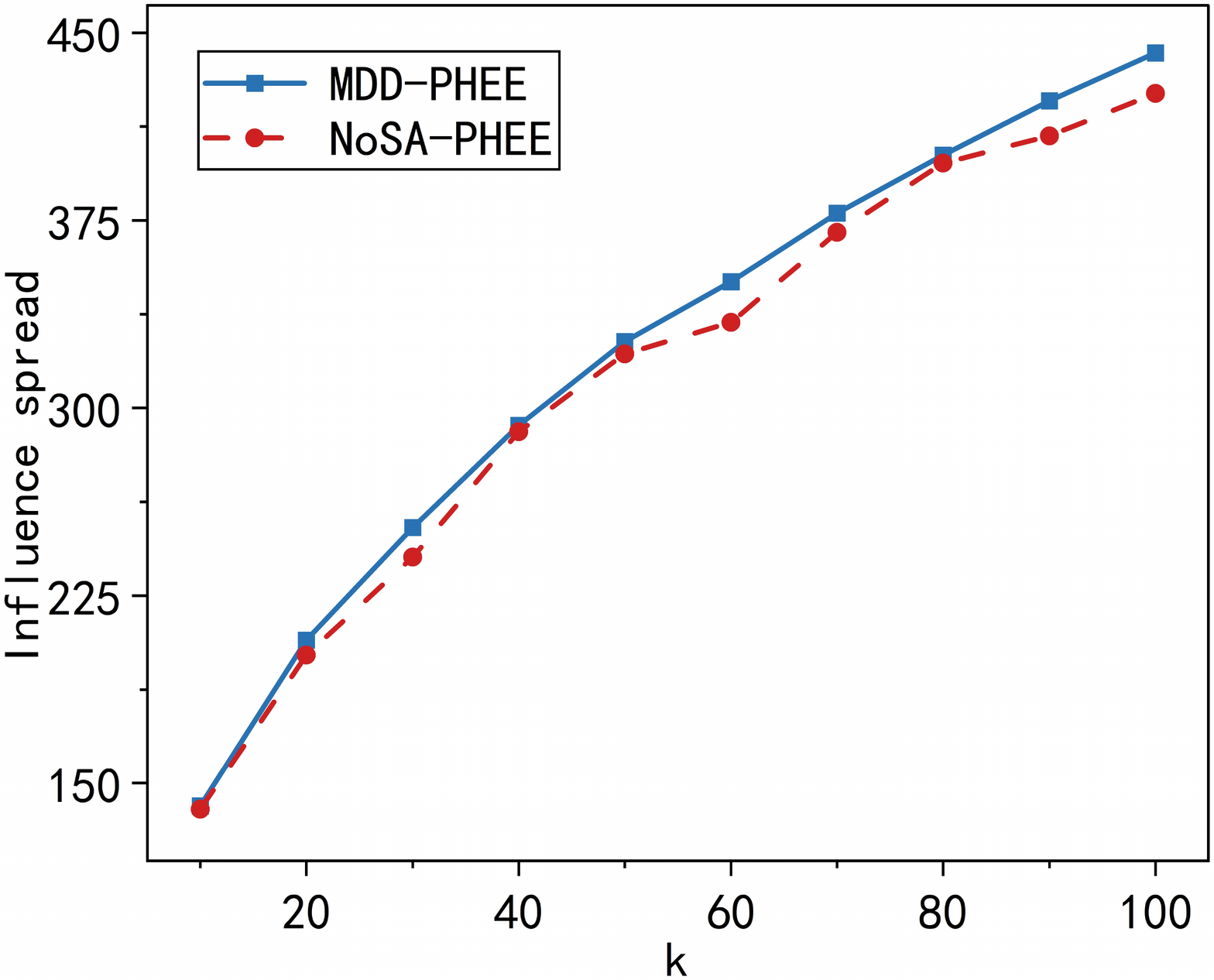}
}
\subfigure[Slashdot]{
\includegraphics[width=3cm]{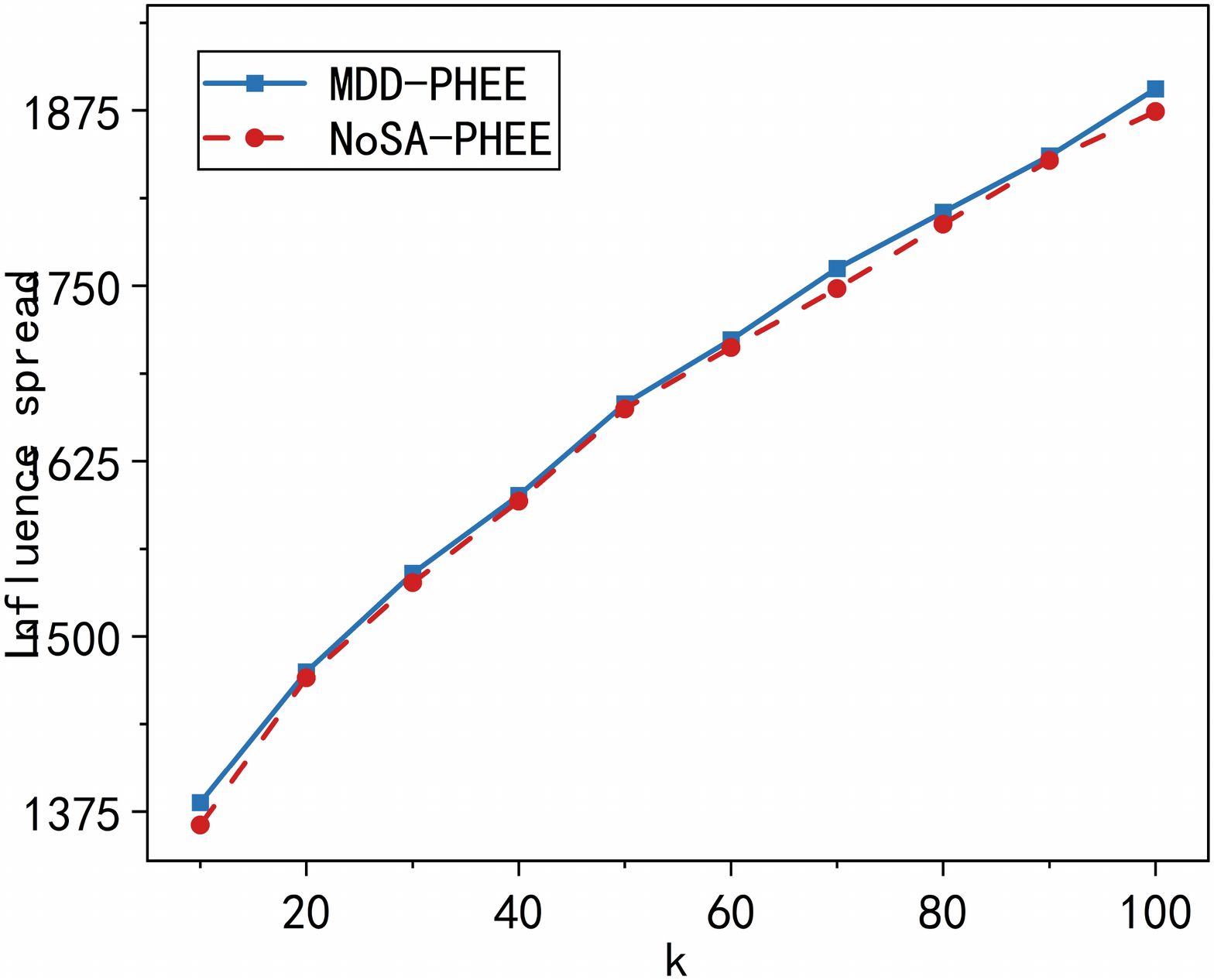}
}
\subfigure[Email-eu]{
\includegraphics[width=3cm]{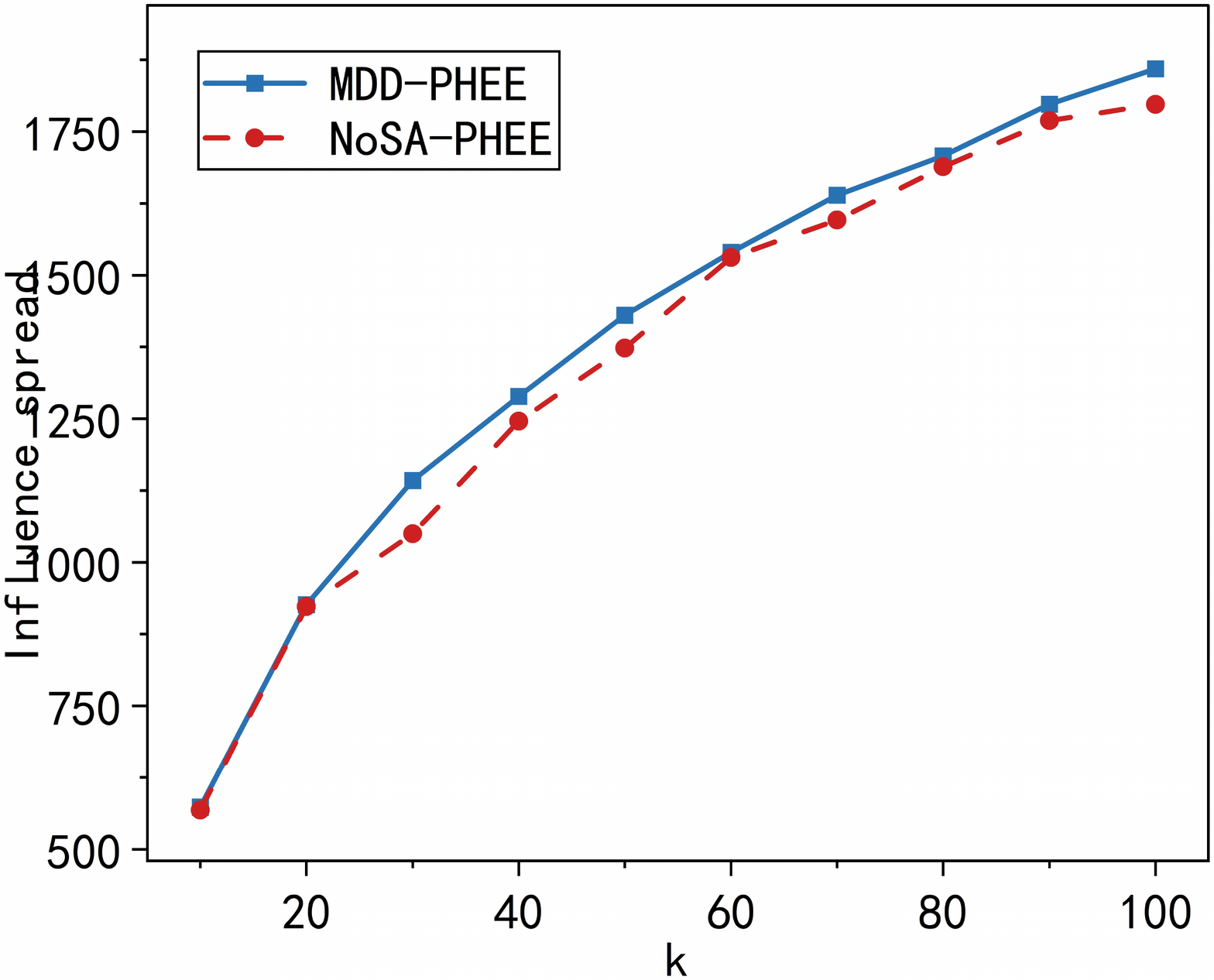}
}
\caption{The comparison on influence spread of MDD-PHEE and NoSA-PHEE.}
\label{fig-exp-4}
\end{center}
\end{figure}

The PHEE framework uses an evolutionary algorithm to enhance the evaluation of vertex influence, and simulated annealing to accelerate convergence. The comparison results between PHEE and SAW-ASA  (section \ref{com-influenc}) have demonstrated the validity of our evolutionary strategy RandRDE(). We conducted a self-comparison experiment to show the effectiveness of the simulated annealing strategy. We compared MDD-PHEE with a modified algorithm, NoSA-PHEE, obtained by removing SAW-ASA from MDD-PHEE and utilizing a neighborhood-based optimization-local search  \cite{gong2016influence}) to improve the individual (seed set) with the maximum EDV value among the population generated by  $RandRDE()$, with   population size $pop=10$,   number of iterations $gmax=150$,    diversity factor $div\_factor$=0.6,  mutation probability $mp=0.1$, and   crossover probability $cp=0.6.$  We ran MDD-PHEE and NoSA-PHEE with seed set sizes from 10
to 100 on the 10 networks under the IC model. As illustrated in Figure \ref{fig-exp-4}, MDD-PHEE clearly outperformed NoSA-PHEE on  eight networks, and NoSA-PHEE can achieve an influence spread close to MDD-PHEE only on Gnutella and Slashdot.

\subsection{Statistical tests}
We carry out a nonparametric statistical analysis based on the influence spread results in section \ref{com-influenc},     employing  two commonly applied methods \cite{biswas2021}, the Friedman test \cite{friedman1937use} and Wilcoxon's signed rank test \cite{Gehan1965}. From the Friedman test, we obtained the performance ranks of all compared algorithms, and from Wilcoxon's signed rank test, we got the performance gaps between pairs of algorithms.

On each network, the Friedman test ranks algorithms according to their influence spreads at each seed set size, and we calculated the mean ranks over all seed set sizes as the final rank of an algorithm on a dataset, as  shown in Table \ref{ftest} (columns 2-11). The overall rank of each algorithm is determined by the average of
all mean ranks; see the last column in Table \ref{ftest}. Again,  an algorithm with better influence spread is reflected by a higher mean rank value.
As shown in Table \ref{ftest}, MDD-PHEE had the highest rank value of 5.225, followed by CELF, GCI-PHEE, SAW-ASA, LIDDE, and DDSE, in decreasing order of rank value. Therefore,  MDD-PHEE outperformed the other algorithms, and GCI-PHEE placed third.

To further show the superior performance of MDD-PHEE, we performed Wilcoxon's signed rank test \cite{Gehan1965} to investigate the difference between it and four top-ranked algorithms in the Friedman test, i.e., CELF, GCI-PHEE, SAW-ASA, and LIDDE.  Table \ref{wtest} describes the results of Wilcoxon's signed rank test, with a significance level of 0.05, where the signs ``+" and ``-" indicate that an algorithm is significantly better or worse, respectively, than another, and ``$\approx$" indicates no significant difference (at P-value $>0.05$). As shown in Table \ref{wtest}, MDD-PHEE outperforms CELF  on small networks, and performs almost the same on large networks, except for CA-AstroPh and Email-eu.
MDD-PHEE outperforms SAW-ASA and LIDDE on almost all networks, and performs similarly to SAW-ASA on CA-GrQc  and to LIDDE on Email-un and soc-Epinions1. Finally, since the performance of GCI  is not very stable on distinct networks, GCI-PHEE performs similarly to MDD-PHEE only on CA-GrQc, CA-HepTh, Gnutella, and soc-Epinions2, and is significantly worse on all other networks.

\begin{table*} [htp!]
\tiny
\centering
\caption{Friedman test results for ranking compared algorithms based on their average influence spreads}
\label{ftest}
\renewcommand\tabcolsep{4.0pt}
\renewcommand{\arraystretch}{1.2}
\begin{tabular}{c c c c c c c c c c c c}
\hline\\
\multirow{2}{*}{Algorithm} & \multicolumn{10}{c}{ Mean rank on each dataset}& \multirow{2}{*}{Overall rank}\\
\cline{2-11}\\
      &  Netscience &  Email-un &  CA-GrQc & CA-HepTh & CA-AstroPh & Gnutella & soc-Epinions1 & soc-Epinions2 & Slashdot & Email-eu \\

    \hline

 $CELF$ & 3.2 & 3.4& 3.3 &3.2& 5.8 & 5& 3.8 & 4.9 &5.1&   5.5 & 4.32\\
$DDSE$  & 1.4& 1.7& 2.8& 2.6& 2.1& 1.9& 2& 1.4& 1.1&      1.2& 1.82\\
$SAW-ASA$& 4.2 & 2.6& 3.8& 4.1& 3.1& 1.4& 1.7& 2.3& 2.9&  1.8& 2.79\\
$MDD-PHEE$& 5.8& 5.5& 4.6& 5.6& 5.2& 5.25& 5.4& 5& 5.1&   4.8& 5.225\\
$GCI-PHEE$& 4.3& 4.1& 5.4& 4.5& 3.8& 4.75& 2.9& 3.7& 4.6& 3.6& 4.165\\
$LIDDE$ & 2.1& 3.7& 1.1& 1& 1& 2.7& 5.1& 3.7& 2.2&       4.1& 2.67\\
    \hline
\end{tabular}
\end{table*}

\begin{table*} [htp!]
\tiny
\centering
\caption{Wilcoxon signed ranks test results for comparing MDD-PHEE with other four competitive algorithms}
\label{wtest}
\renewcommand\tabcolsep{3.0pt}
\renewcommand{\arraystretch}{1.2}
\begin{tabular}{c c c c c| c c c c| c c c c| c c c c}
\hline\\
\multirow{3}{*}{Dataset} & \multicolumn{16}{c}{Algorithm comparison}\\
\cline{2-17}\\
& \multicolumn{4}{c|}{MDD-PHEE $vs$ CELF} & \multicolumn{4}{c|}{MDD-PHEE $vs$ SAW-ASA} & \multicolumn{4}{c|}{MDD-PHEE $vs$ LIDDE} & \multicolumn{4}{c}{MDD-PHEE $vs$ GCI-PHEE} \\
\cline{2-5}  \cline{6-9} \cline{10-13}  \cline{14-17}  \\

      &  Better & Worse & P-value & Decision&     Better & Worse & P-value & Decision&    Better & Worse & P-value & Decision&    Better & Worse & P-value & Decision\\

    \hline
NetScience & 10 &0 & 0.005& +  & 10 &0 & 0.005& + & 10 &0 & 0.005& + & 9 &1 & 0.007& +
\\

Email-un & 9 &1 & 0.009&  + & 10 &0 & 0.005& + & 7 &3 & 0.059& $\approx$ & 9 &1 & 0.017& +
\\

CA-GrQc  & 7 &3 & 0.028&  + & 7 &3 & 0.093& $\approx$ & 10 &0 & 0.005& + & 3 &7 & 0.241& $\approx$
\\

CA-HetPh  & 9 &1 & 0.007&  + & 10 &0 & 0.005& + & 10 &0 & 0.005& + & 7 &3 & 0.139& $\approx$
\\

CA-AstroPh & 2 &8 & 0.037&  - & 10 &0 & 0.005& + & 10 &0 & 0.005& +
 & 10 &0 & 0.005& + \\

Gnutell & 5 &5 & 0.508&  $\approx$ & 10 &0 & 0.005& + & 10 &0 & 0.005& + & 6 &4 & 0.066& $\approx$
\\

soc-Epinions1 & 8 &2 & 0.093&  $\approx$ & 10 &0 & 0.005& + & 7 &3 & 0.878& $\approx$ & 9 &1 & 0.007& +
\\

soc-Epinions2 & 4 &6 & 0.093&  $\approx$ & 9 &1 & 0.007& + & 10 &0 & 0.005& +& 7 &3 & 0.074& $\approx$
\\

Slashdot & 5 &5 & 0.959&  $\approx$ & 10 &0 & 0.005& + & 8 &2 & 0.013& + & 9 &1 & 0.007& +
\\

Email-eu & 2 &8 & 0.047& - & 10 &0 & 0.005& + & 8 &2 & 0.013& + & 8 &2 & 0.013& +
\\
\hline
\end{tabular}
\end{table*}

\section{Conclusion and future work} \label{sec7}
In this study, we proposed PHEE for the IM problem, based on which two algorithms MDD-PHEE and GCI-PHEE are developed. We changed the conventional paradigm of evolutionary approaches by adding simulated annealing to accelerate convergence. This improved the flexibility of the algorithm to cope with the risk of become trapped in a local optimum. Random range division improved solution diversity. Experimental results on 10 real networks of different sizes and structures showed the superiority of the PHEE framework to four baseline algorithms.

Although both GCI and MDD were based on the $k$-shell, GCI-PHEE performs worse than MDD-PHEE, perhaps due to the intrinsic characteristic of GCI, which
was influenced by the network structure. Nevertheless, GCI-PHEE was competitive with the compared algorithms, which also showed the effectiveness of PHEE. We attempted to replace MDD or GCI with $K$-shell, but without good results. Therefore,  the vertex ranking method was found to influence the performance of PHEE. Our future research directions include investigating a more suitable vertex ranking strategy.

We explored  other evolutionary approaches for the first stage of PHEE, to realize a better performance than the current two PHEE algorithms. We experimented with a differential evolutionary approach, but it failed to balance search efficiency and solution accuracy. To explore a more suitable evolutionary approach for PHEE is another  future research direction. Finally, possible applications of PHEE to other optimization problems should also be explored.

\section*{Acknowledgment}

This work was supported in part by the National Natural Science Foundation of China
under Grants (61872101, 62172072), in part by Natural Science Foundation of Guangdong Province of China under Grant 2021A1515011940.

\section*{Reference}

\bibliographystyle{IEEEtran}

\bibliography{mybib}

\end{document}